%% file: main.tex
\pdfoutput=1
\documentclass{article}
\RequirePackage[l2tabu, orthodox]{nag}

\usepackage{graphicx} 
\usepackage{mathtools}
\usepackage{braket}
\usepackage{amsmath}
\usepackage{amsfonts}
\usepackage{subcaption}
\usepackage{pgfplots} 
\usepgfplotslibrary{fillbetween}
\pgfplotsset{compat=newest}
\usepackage{framed}
\usepackage{dsfont}
\usepackage{tikz}
\usepackage{tikz-3dplot} 
\usetikzlibrary{decorations,shapes,chains,fit,intersections,arrows,calc,positioning,shapes.geometric, decorations.markings}
\usepackage[affil-it]{authblk}
\usepackage{rotating}
\usepackage{float}
\usepackage{hhline}
\usepackage{tablefootnote}
\usepackage{booktabs}
\usepackage{titling}

\usepackage[margin=2cm]{geometry}

\usepackage{xcolor}
\usepackage{xstring}
\usepackage{xcolor}
\definecolor{dullmagenta}{rgb}{0.5,0,0.4}
\definecolor{dullblue}{rgb}{0.0,0,0.86}
\usepackage[hyperfootnotes=false]{hyperref}
\hypersetup{colorlinks=true,citecolor=dullblue,linkcolor=dullblue,filecolor=dullblue,urlcolor=dullblue,breaklinks=true,bookmarksnumbered=true, bookmarksopen=true,bookmarksopenlevel=1}
\usepackage[capitalise]{cleveref}
\usepackage{adjustbox}

\DeclareFontFamily{U}{bbold}{}
\DeclareFontShape{U}{bbold}{m}{n}
 {
  <-5.5> s*[1.069] bbold5
  <5.5-6.5> s*[1.069] bbold6
  <6.5-7.5> s*[1.069] bbold7
  <7.5-8.5> s*[1.069] bbold8
  <8.5-9.5> s*[1.069] bbold9
  <9.5-11> s*[1.069] bbold12 
  <11-15> s*[1.069] bbold12
  <15-> s*[1.069] bbold17
 }{}

\DeclareRobustCommand{\id}{%
  \text{\usefont{U}{bbold}{m}{n}1}%
}

\DeclareMathOperator*{\argmax}{argmax}
\newcommand{\pgroup}[1]{\mathcal{P}_{#1}}
\newcommand{\ppgroup}[1]{\mathcal{P}^*_{#1}}
\newcommand{\plog}{\mathcal{L}^*}
\newcommand{\pstab}{\mathcal{S}^*}
\newcommand{\pdestab}{\mathcal{D}^*}

\def\generator{generator}
\def\eq{\vphantom{$+$}$=$}
\tikzset{
    max width/.style args={#1}{
        execute at begin node={\varwidth{#1}},
        execute at end node={\endvarwidth}
    }
}

\definecolor{cmap3}{rgb}{0.10588235294117647, 0.6196078431372549, 0.4666666666666667}
\definecolor{cmap5}{rgb}{0.8509803921568627, 0.37254901960784315, 0.00784313725490196}
\definecolor{cmap7}{rgb}{0.4588235294117647, 0.4392156862745098, 0.7019607843137254}
\definecolor{cmap9}{rgb}{0.9058823529411765, 0.1607843137254902, 0.5411764705882353}
\definecolor{cmap11}{rgb}{0.4, 0.6509803921568628, 0.11764705882352941}
\newcommand{\drawline}[3]{ 
\addplot [name path=lower, fill=none, draw=none] table [x=X, y expr=\thisrow{Y#1} - \thisrow{YERR#1}, col sep=comma]{#3};
\addplot [name path=upper, fill=none, draw=none] table [x=X, y expr=\thisrow{Y#1} + \thisrow{YERR#1}, col sep=comma]{#3};
\addplot[color=#2!20] fill between[of=lower and upper=0];
\addplot[color=#2,thick] table [x=X, y=Y#1, col sep=comma]{#3};
}
\newcommand{\drawlineref}[3]{ 
\addplot[color=#2, dashed,thick] table [x=X, y=Y#1, col sep=comma]{#3};
}
\newcommand{\addtolegend}[2]{ 
\addlegendimage{no markers,color=#2,very thick}
\addlegendentry{$d=#1$}
}

\newenvironment{aligns}{\subequations \align} {\endalign \endsubequations}

\usepackage[backend=biber,sorting=none,style=phys-jmr,biblabel=brackets,backref=true,bibencoding=utf8,maxnames=20,eprint=true]{biblatex}
\makeatletter
\pretocmd{\blx@head@bibintoc}{\phantomsection}{}{\ddt}
\makeatother
\DefineBibliographyStrings{english}{%
  backrefpage = {page},
  backrefpages = {pages},
}
\usepackage{multirow}

\title{Tensor Network Decoding Beyond 2D}
\author{Christophe Piveteau, Christopher T.\ Chubb, and Joseph M.\ Renes\vspace{-2ex}}
\affil{\normalsize Institute for Theoretical Physics, ETH Z\"urich, Switzerland}
\date{}

\begin{document}

\maketitle
\vspace{-1em}
\vspace{-2\baselineskip}
\renewcommand{\abstractname}{\vspace{-\baselineskip}}
\vspace{-1.5em}
\begin{abstract}
Decoding algorithms based on approximate tensor network contraction have proven tremendously successful in decoding 2D local quantum codes such as surface/toric codes and color codes, effectively achieving optimal decoding accuracy.
In this work, we introduce several techniques to generalize tensor network decoding to higher dimensions so that it can be applied to 3D codes as well as 2D codes with noisy syndrome measurements (phenomenological noise or circuit-level noise).
    The three-dimensional case is significantly more challenging than 2D, as the involved approximate tensor contraction is dramatically less well-behaved than its 2D counterpart.
    Nonetheless, we numerically demonstrate that the decoding accuracy of our approach outperforms state-of-the-art decoders on the 3D surface code, both in the point and loop sectors, as well as for depolarizing noise. 
    Our techniques could prove useful in near-term experimental demonstrations of quantum error correction, when decoding is to be performed offline and accuracy is of utmost importance.
    To this end, we show how tensor network decoding can be applied to circuit-level noise and demonstrate that it outperforms the matching decoder on the rotated surface code.
    Our code is available at ~\href{https://github.com/ChriPiv/tndecoder3d}{https://github.com/ChriPiv/tndecoder3d}.
\end{abstract}
\vspace{-2em}
\section{Introduction}

The practical realization of quantum technologies is impeded by the inherent sensitivity of quantum systems to noise.
Quantum error correction~\cite{shor_scheme_1995,steane_error_1996} is generally considered the solution to this problem, as it allows for fault-tolerant processing and transmission of quantum information.
The idea is to embed the quantum information into a larger system using a quantum code to increase its resilience against noise.
The stabilizer formalism~\cite{gottesman_stabilizer_1997} has proven to be a convenient and useful framework to construct and describe quantum codes.
The error correction procedure for a stabilizer code involves the measurement of certain stabilizer operators, and the measurement outcomes (the syndrome) are subsequently processed by a classical decoding algorithm (or decoder) to select a correction operation to be applied on the system.
Generally speaking, decoders are evaluated according to two criteria: speed and accuracy.
A low latency will be crucial for long-term hardware implementations, as the errors must be decoded and corrected faster than they can build up.
There is also significant interest in studying slower, but more accurate decoders.
They are crucial tools to study and benchmark fundamental properties of quantum codes and other less accurate decoders, and have previously been used to aid the design of quantum codes~\cite{delfosse_linear-time_2016}.
They also lend themselves to experimental demonstrations of error-corrected quantum memories on near-term hardware, when the decoding process is performed offline and runtime is not an immediate issue~\cite{acharya2023_suppressing}.

Tensor network (TN) decoders operate by approximately contracting one or more relevant tensor networks and subsequently deciding on the correction operation from the contraction results. 
The relevant tensor networks describe the probability of a given logical error conditioned on the observed syndrome. 
If the tensor networks were contracted exactly, the resulting decoder would be optimal, so the only heuristic lies in the approximate contraction.
Generally, the layout of the networks follows the locality of the code from which they were derived, and in the case of correlated noise from the locality of those correlations~\cite{chubb_statistical_2021}.
Previous work has focused mainly on 2D local codes, i.e.\ codes whose stabilizer operators are local when the qubits are laid out in a two-dimensional array.
This family of codes includes topological codes, such as the 2D toric/surface code~\cite{kitaev_fault-tolerant_2003} and the 2D color code~\cite{bombin_topological_2006}.
TN decoding performs remarkably well on all of these codes and has been shown to achieve essentially optimal decoding accuracy at competitive runtimes for bit-flip, phase-flip, and depolarizing noise~\cite{bravyi_efficient_2014,chubb_statistical_2021,chubb_general_2021}. 
Another approach is to base the tensor network on the encoding circuit of the code~\cite{ferrisTensorNetworksQuantum2014}; this has recently been shown to perform well on codes with local circuits of low depth~\cite{darmawan_low-depth_2022}.

However, while approximate contraction of 2D TNs is well understood and in most cases numerically well-behaved, approximate contraction of 3D networks is not and  remains an active field of research.
The fundamental obstruction is that it is not possible to define a canonical gauge of a tensor network that is not a tree~\cite[\S5.2]{orus_practical_2014}, so for example optimally truncating a projected entangled pair state (PEPS) network is more difficult than a matrix product state (MPS).
This restriction is especially unfortunate for practical applications, since in experimental platforms the syndrome measurements are noisy and must be repeated multiple times (typically on the order of the distance of the code).
The resulting decoding problem for a 2D local code under circuit-level noise is fundamentally three-dimensional, and thus involves three-dimensional TNs.
For this reason, existing TN decoding schemes are not applicable.

Here, we introduce a family of techniques which extend the framework of TN decoding beyond two dimensions.
The resulting decoding accuracy can in principle be made arbitrarily close to optimal, and thus we focus on accuracy and eschew the issue of runtime. 
Decoding the 2D surface code with noisy syndromes and repeated measurements rounds is mathematically equivalent to decoding the 3D surface code~\cite{dennis_topological_2002}, so our work finally makes it possible to probe these problems with a near-optimal decoder.

\bigskip

We describe our main contributions in more detail below, by way of sketching the structure of the remainder of the paper. 
The following~\Cref{sec:prelim} reviews the properties of stabilizer codes, their optimal decoding as finding the most probable logical coset consistent with the syndrome, and the formalism of detector error models. It also gives a short overview of tensor networks. 

In~\Cref{sec:optimaldecoding} we introduce two distinct representations (applicable to any stabilizer code) of the logical coset probabilities as TN contractions. 
The \emph{generator picture} mathematically corresponds to the statistical-mechanical mapping used in \cite{bravyi_efficient_2014,chubb_statistical_2021,chubb_general_2021}, but without the associated statistical-mechanical language.
The \emph{detector picture} can be regarded as a dual representation to the generator picture. 
To our knowledge it has not been previously proposed in the literature in this context.
The idea of the detector picture is to choose the tensor network to represent a sum over all possible error patterns, and then remove the patterns that are not compatible with an observed syndrome and a given logical operator.
In contrast, the generator picture takes the opposite approach, first fixing a representative Pauli error $r$ that is compatible with the observed syndrome and logical operator, and then summing over the probabilities of errors $r\cdot s$ over all stabilizers $s$ to obtain the total probability of the corresponding logical coset.

In~\Cref{sec:approxcontract} we propose a technique for approximate contraction of both detector and generator TNs: Sweep a two-dimensional TN from one side of the 3D TN to the other, contracting the tensors that are encountered along the way layer-by-layer, all while continuously truncating the bond dimensions of the 2D network to keep it below a given threshold.
This can be seen as a higher-dimensional generalization of the sweep-line algorithm used in~\cite{chubb_general_2021}.
Unlike the case of sweeping with an MPS, one cannot define a canonical form for a 2D tensor network that includes loops, making it difficult to realize a numerically well-behaved truncation scheme. 
We make use of the \emph{simple update} technique~\cite{jahromi_universal_2019,corboz_simulation_2010,jiang_accurate_2008}, commonly used in the condensed matter community to simulate the time evolution of spin chains. 
The idea is to keep track of a rank-one approximation of the environment of each tensor in the loopy tensor network, which allows for a more accurate truncation.
This method was recently shown to be sufficient for fast and highly accurate simulations of 2D quantum systems with more than a thousand qubits~\cite{patra_efficient_2023}.
Its theoretical justification lies in the fact that it makes the loopy network approach the \emph{Vidal gauge} when the applied gates are close enough to the identity~\cite{tindall_gauging_2023}.

In~\Cref{sec:results} we present the results of numerical simulations of TN decoding.
First, we consider independent bit- and phase-flip noise on the unrotated 3D surface code, differentiating the stabilizers into the \emph{point sector} of weight-six $X$-type stabilizers and the \emph{loop sector} of weight-four $Z$-type stabilizers.
Note that decoding the point sector can equivalently be regarded as decoding the 2D surface code with a phenomenological noise model~\cite{dennis_topological_2002}.
These two sectors are decoded with detector and generator TNs, respectively. 
As shown in~\Cref{fig:threshold_z,fig:threshold_x}, TN decoding outperforms the minimum-weight perfect matching decoder for the point sector~\cite{wangConfinementHiggsTransitionDisordered2003} and BP+OSD for the loop sector.
To our knowledge, BP+OSD is the state of the art for the loop sector~\cite{quintavalle_single-shot_2021}.
In a second step, we consider depolarizing noise on the unrotated 3D surface code, which involves both the point and loop sector in a correlated manner.
We find that the detector TN decoder performs considerably better than the generator TN.
As depicted in~\Cref{fig:threshold_depol}, it also significantly improves over the existing state-of-the-art decoder, which is BP+OSD~\cite{huang_tailoring_2022}.

We also develop a technique for TN decoding of circuit-level noise.
One could make use of the circuit itself to construct the TN, but while the resulting TNs are three-dimensional, they are very large even for modest-sized decoding tasks.
Instead, we make use of Stim's detector error models~\cite{gidney_stim_2021} to construct the TN, and consider the rotated surface code as a proof-of-principle.
Even with this approach, the resulting 3D tensor network is notoriously large and exhibits a very complicated topology, as seen in~\Cref{fig:tn_circ,fig:tn_circ_large}.
To solve this problem, we first ``compress'' the TN to a cubic form via local approximate contraction.
The result can then be globally contracted with the simple update method described above.
Numerical results for the rotated surface code are depicted in~\Cref{fig:threshold_circ} and compared to PyMatching~\cite{higgott_pymatching_2022,higgott_sparse_2023}.
We note that TN decoding of circuit-level noise has been considered in~\cite{bohdanowicz2022quantum} before, but the authors were only able to consider very small codes of distance 3 due to numerical challenges with their technique.

\Cref{tab:thresholds} summarizes our observed thresholds and compares them with decoders in previous literature.
The thresholds are estimated using the bootstrapping procedure explained in~\cite{chubb_general_2021}, see~\cite{chubb_boostrap_2023} for more details.
We note that these thresholds should be taken with a grain of salt, since the considered code distances are rather small and therefore finite-size effects might have a significant effect.
We argue that the improvements in logical error rates are more significant here, especially for the circuit-level noise where the exact threshold might be of small importance for experiments.

Finally, we discuss the ample directions for further research in~\Cref{sec:discussion}. 
Our software implementation of the tensor network decoder is available at~\href{https://github.com/ChriPiv/tndecoder3d}{https://github.com/ChriPiv/tndecoder3d}.

\captionsetup[subfigure]{margin=-100cm, singlelinecheck=false}

\newpage 
\begin{figure}[H]
    \centering
    \begin{tikzpicture}
        \begin{axis}[hide axis,xmin=10,xmax=50,ymin=0,ymax=0.4,legend columns=-1]{}
         \addtolegend{3}{cmap3}
         \addtolegend{5}{cmap5}
         \addtolegend{7}{cmap7}
         \addtolegend{9}{cmap9}
         \addtolegend{11}{cmap11}
        \end{axis}
    \end{tikzpicture}
    \hfill \\
    \vspace{5pt}
    \begin{subfigure}[b]{0.49\textwidth}
        \centering
        \begin{tikzpicture}
        \begin{axis}[
            width=\textwidth,
            height=0.7\textwidth,
            ylabel=logical error rate,
            legend pos=north west,
            xtick = {0.029, 0.030, 0.031, 0.032, 0.033},
            xticklabels = {2.9\%, 3.0\%, 3.1\%, 3.2\%, 3.3\%},
            ytick = {0.06,0.07,0.08,0.09,0.10,0.11},
            yticklabels = {6\%, 7\%, 8\%, 9\%, 10\%, 11\%},
            name=border,
            scaled x ticks=false
        ]
        
        \drawline{3}{cmap3}{data/data_Z_tn.txt}
        \drawline{5}{cmap5}{data/data_Z_tn.txt}
        \drawline{7}{cmap7}{data/data_Z_tn.txt}
        \drawline{9}{cmap9}{data/data_Z_tn.txt}
        \drawline{11}{cmap11}{data/data_Z_tn.txt}
    
        \drawlineref{3}{cmap3}{data/data_Z_matching.txt}
        \drawlineref{5}{cmap5}{data/data_Z_matching.txt}
        \drawlineref{7}{cmap7}{data/data_Z_matching.txt}
        \drawlineref{9}{cmap9}{data/data_Z_matching.txt}
        \drawlineref{11}{cmap11}{data/data_Z_matching.txt}
    
        \node[] at (axis cs: 0.0295,0.095) {MWPM};
        \node[] at (axis cs: 0.032,0.065) {TN decoder};
    
        \end{axis}
        \path (border.north west) ++(1, -0.8) node[above left]{(a)};
        \end{tikzpicture}
        \caption{}
        \label{fig:threshold_z}
    \end{subfigure}
    \hfill
    \begin{subfigure}[b]{0.49\textwidth}
        \centering
        \begin{tikzpicture}
        \begin{axis}[
            width=\textwidth,
            height=0.7\textwidth,
            legend pos=north west,
            xtick={0.2, 0.21, 0.22, 0.23},
            xticklabels={20\%, 21\%, 22\%, 23\%},
            ytick={0.1, 0.15, 0.2, 0.25, 0.3, 0.35},
            yticklabels={10\%, 15\%, 20\%, 25\%, 30\%, 35\%},
            name=border
        ]
        
        \drawline{3}{cmap3}{data/data_X_tn.txt}
        \drawline{5}{cmap5}{data/data_X_tn.txt}
        \drawline{7}{cmap7}{data/data_X_tn.txt}
        \drawline{9}{cmap9}{data/data_X_tn.txt}
    
        \drawlineref{3}{cmap3}{data/data_X_bposd.txt}
        \drawlineref{5}{cmap5}{data/data_X_bposd.txt}
        \drawlineref{7}{cmap7}{data/data_X_bposd.txt}
        \drawlineref{9}{cmap9}{data/data_X_bposd.txt}

        \node[] at (axis cs: 0.21,0.28) {BP+OSD};
        \node[] at (axis cs: 0.22,0.14) {TN decoder};
        \end{axis}
        \path (border.north west) ++(1, -0.8) node[above left]{(b)};
        \end{tikzpicture}
        \caption{}
        \label{fig:threshold_x}
    \end{subfigure}
    \hfill \\
    \vspace{-15pt}
    \begin{subfigure}[b]{0.49\textwidth}
        \centering
        \begin{tikzpicture}
        \begin{axis}[
            width=\textwidth,
            height=0.7\textwidth,
            xlabel=physical error rate,
            ylabel=logical error rate,
            legend pos=north west,
            xtick={0.062, 0.066, 0.070, 0.074},
            xticklabels={6.2\%, 6.6\%, 7.0\%, 7.4\%},
            scaled x ticks=false,
            ytick = {0.05,0.07,0.09,0.11,0.13,0.15},
            yticklabels = {5\%, 7\%, 9\%, 11\%, 13\%, 15\%},
            name=border
]
        
        \drawline{3}{cmap3}{data/data_depol_tn.txt}
        \drawline{5}{cmap5}{data/data_depol_tn.txt}
        \drawline{7}{cmap7}{data/data_depol_tn.txt}
        \drawline{9}{cmap9}{data/data_depol_tn.txt}
        \drawline{11}{cmap11}{data/data_depol_tn.txt}
    
        \drawlineref{3}{cmap3}{data/data_depol_bposd.txt}
        \drawlineref{5}{cmap5}{data/data_depol_bposd.txt}
        \drawlineref{7}{cmap7}{data/data_depol_bposd.txt}
        \drawlineref{9}{cmap9}{data/data_depol_bposd.txt}
        \drawlineref{11}{cmap11}{data/data_depol_bposd.txt}
    
        \node[] at (axis cs: 0.066,0.11) {BP+OSD};
        \node[] at (axis cs: 0.072,0.06) {TN decoder};
    
        \end{axis}
        \path (border.north west) ++(1, -0.8) node[above left]{(c)};
        \end{tikzpicture}
        \caption{}
        \label{fig:threshold_depol}
    \end{subfigure}
    \hfill
    \begin{subfigure}[b]{0.49\textwidth}
        \centering
        \begin{tikzpicture}
        \begin{axis}[
            width=\textwidth,
            height=0.7\textwidth,
            xlabel=physical error rate,
            legend pos=north west,
            xtick={0.006, 0.007, 0.008, 0.009},
            xticklabels={0.6\%, 0.7\%, 0.8\%, 0.9\%},
            scaled x ticks=false,
            ytick={0.01, 0.02, 0.03, 0.04},
            yticklabels={1\%, 2\%, 3\%, 4\%},
            scaled y ticks=false,
            name=border
        ]
        
        \drawline{7}{cmap7}{data/data_circ_tn.txt}
        \drawline{5}{cmap5}{data/data_circ_tn.txt}
        \drawline{3}{cmap3}{data/data_circ_tn.txt}
        
        \drawlineref{3}{cmap3}{data/data_circ_matching.txt}
        \drawlineref{5}{cmap5}{data/data_circ_matching.txt}
        \drawlineref{7}{cmap7}{data/data_circ_matching.txt}
    
        \node[] at (axis cs: 0.0070,0.032) {MWPM};
        \node[] at (axis cs: 0.0085,0.015) {TN decoder};
    
        \end{axis}
        \path (border.north west) ++(1, -0.8) node[above left]{(d)};
        \end{tikzpicture}
        \caption{}
        \label{fig:threshold_circ}
    \end{subfigure}
    \caption{Threshold plots for (a) point sector, (b) loop sector, (c) depolarizing noise for the 3D unrotated surface code. We compare the tensor network decoder with minimum-weight perfect matching (a) and the BP+OSD decoder (b,c), which is to our knowledge the state of the art decoders for these two sectors. (d) depicts how the tensor network decoder compares to the matching decoder for circuit-level noise on the rotated surface code. The thresholds are depicted in~\Cref{tab:thresholds}.}
    \label{fig:three graphs}
\end{figure}
\captionsetup[subfigure]{margin=0pt, singlelinecheck=true}

\begin{table}[H]
\centering
\begin{tabular}{ccrlc}
\toprule
 & \textbf{TN decoder (ours)} & \multicolumn{2}{c}{\textbf{Other decoders}} & \textbf{Optimal threshold} \\ \hline
\multirow{2}{*}{\textbf{Point sector}} & \multirow{2}{*}{$3.136^{+0.012}_{-0.014}\%$} & Matching~\cite{wangConfinementHiggsTransitionDisordered2003} & $2.93\pm .02\%$ & \multirow{2}{*}{$\approx 3.3\%$~\cite{ohno_phase_2004}} \\
 &  & RG~\cite{dulcoscianci_faulttolerant_2014} & $1.9\pm 0.4 \%$ &  \\ \hline
\multirow{6}{*}{\textbf{Loop sector}} & \multirow{6}{*}{$22.788^{+0.123}_{-0.107}\%$} & Erasure mapping~\cite{aloshious_decoding_2021} & $\approx 12.2\%$ & \multirow{6}{*}{\begin{tabular}[c]{@{}c@{}}$23.180\pm 0.004\%$\\ \cite{ozeki_multicritical_1998,hasenbusch_magnetic-glassy_2007}\end{tabular}} \\
 &  & Toom's rule~\cite{kubica_the_2017} & $\approx 14.5\%$ &  \\
 &  & Sweep~\cite{vasmer_cellular_2021} & $\approx 15.5\%$ &  \\
 &  & RG~\cite{duivenvoorden_renormalization_2019} & $\approx 17.2\%$ &  \\
 &  & Neural network~\cite{breuckmann_scalable_2018} & $\approx 17.5\%$ &  \\
 &  & BP+OSD~\cite{quintavalle_single-shot_2021} & $21.55\pm 0.01\%$ &  \\ \hline
\multirow{2}{*}{\textbf{Depolarizing}} & \multirow{2}{*}{$7.067^{+0.034}_{-0.033}\%$} & BP+OSD~\cite{huang_tailoring_2022} & $5.95\pm0.03\%$ & \multirow{2}{*}{unknown} \\
 &  & BP+OSD (ours)\footnotemark & $6.715\pm0.012\%$ &  \\ \hline
\textbf{\begin{tabular}[c]{@{}c@{}}Circuit-level\\ depolarizing\end{tabular}} & $\approx 0.8\%$ & Matching (ours) & $\approx 0.78\%$ & unknown \\ \bottomrule
\end{tabular}
    \caption{Comparison of thresholds between TN decoders and prior art. In some cases, the optimal threshold can be estimated using a statistical mechanical mapping. Loop sector references adapted from~\cite{quintavalle_single-shot_2021}.
    Entries with ``(ours)'' stem from numerics described in this manuscript.}
    \label{tab:thresholds}
\end{table}
\footnotetext[1]{There might be some technical differences between the two BP+OSD implementations. Furthermore we do not consider periodic boundary conditions and the small distances might lead to finite size effects playing a significant role.}
\newpage

\section{Preliminaries}
\label{sec:prelim}
\subsection{The Pauli Group and its Symplectic Structure}
The $n$-qubit Pauli group $\pgroup{n}$ is defined as the set of $n$-fold tensor products of single-qubit Pauli operators with an additional phase $\pm 1$ or $\pm i$:
\begin{equation}
    \pgroup{n} := \left\{ i^c\cdot Q^{(1)}\otimes\dots\otimes Q^{(n)} \,\middle|\, c\in\{0,1,2,3\}, Q^{(i)}\in\{\mathds{1},\sigma_X,\sigma_Y,\sigma_Z\} \right\}
\end{equation}
where $\sigma_X, \sigma_Y$ and $\sigma_Z$ denote the three Pauli matrices.
For $Q\in\pgroup{n}$ we denote its $i$-th component by $Q^{(i)}\in\{\mathds{1},\sigma_X,\sigma_Y,\sigma_Z\}$.
Any Pauli operator can be represented (up to phase) by a pair of length-$n$ binary bitstrings $u,v\in \mathbb F_2^n$ defined as follows:
\begin{equation}
    u_i := \begin{cases}1 &\text{ if } Q^{(i)}\in \{\sigma_X,\sigma_Y\} \\ 0 &\text{ else}\end{cases} \quad \, \quad v_i := \begin{cases}1 &\text{ if } Q^{(i)}\in \{\sigma_Z,\sigma_Y\} \\ 0 &\text{ else}\end{cases}
\end{equation}
The map $w:\pgroup{n}\to \mathbb F_2^{2n}$ taking $Q$ to $(u,v)$ is a group homomorphism in that $w(Q'Q)=w(Q')+w(Q)$. 

The fact that any two Pauli operators either commute or anticommute allows us to define a symplectic form on $\pgroup{n}$, namely $\langle \cdot,\cdot\rangle: \pgroup{n}\times \pgroup{n}\to \mathbb F_2$ with 
\begin{equation}
\langle Q,Q'\rangle := \begin{cases} 0 &\text{ if } [Q,Q']=0,  \\ 1 &\text{ if } \{Q,Q'\}=0. \end{cases}
\end{equation}
Observe that $\langle Q,Q'\rangle=w(Q)^TJw(Q')$ for $J=\left(\begin{smallmatrix} 0 & \id\\ \id & 0\end{smallmatrix}\right)$, when interpreting $w(Q)$ as a column vector, with $\id$ the $n$-dimensional identity matrix. 

A symplectic basis for $\mathcal P_n$ is a set of $2n$ Pauli operators $X_1,\dots,X_n,Z_1,\dots,Z_n \in \mathcal{P}_n$ which fulfills the (anti-)commutation relations
\begin{equation}
    \langle X_i,X_j\rangle = 0\,,\qquad 
    \langle Z_i,Z_j\rangle = 0\,,\qquad 
    \text{and}\qquad
    \langle X_i,Z_j\rangle = \delta_{ij}\
\end{equation}
for all $i,j\in\{1,\dots,n\}$.
Given a symplectic basis, any Pauli operator $Q\in\pgroup{n}$ can be written as a product of the basis elements
\begin{equation}\label{eq:sympl_basis_decomp}
    Q = i^{c}\cdot X_1^{\lambda_1}\cdots X_n^{\lambda_n}\cdot Z_1^{\mu_1}\cdots Z_n^{\mu_n}
\end{equation}
for some set of coefficients $c\in\{0,1,2,3\}, \lambda,{\mu}\in\mathbb F_2^{2n}$.

In many instances, the phase of a Pauli operator is irrelevant and to reflect this we define the \emph{projective Pauli group} as
\begin{equation}
    \ppgroup{n} := \pgroup{n} / \{\mathds{1}, -\mathds{1}, i\mathds{1}, -i\mathds{1}\}
\end{equation}
which represents the Pauli operators modulo phase.
We denote the projection from $\pgroup{n}\to\ppgroup{n}$ by $\pi$.
For convenience, we will generally denote elements of $\pgroup{n}$ with upper-case letters and elements of $\ppgroup{n}$ with lower-case letters.
Any element $q\in\ppgroup{n}$ can be written as $q=\pi(Q)$ for some $Q\in\pgroup{n}$, so by~\Cref{eq:sympl_basis_decomp} one can write
\begin{equation}
    q = x_1^{\lambda_1}\cdots x_n^{\lambda_n}\cdot z_1^{\mu_1}\cdots z_n^{\mu_n}
\end{equation}
where $x_i:=\pi(X_i)$ and $z_i:=\pi(Z_i)$.
For a given symplectic basis, there is a one-to-one relation between the $4^n$ elements of $\ppgroup{n}$ and the $4^n$ possible coefficients $(\lambda,{\mu})\in\mathbb F_2^{2n}$.

\subsection{Stabilizer Codes}
Here, we sketch the relevant aspects of stabilizer codes; for more details see~\cite{gottesman_introduction_2010}. 
An $n$-qubit stabilizer code is characterized by its \emph{stabilizer group} $\mathcal{S}$, an Abelian subgroup of $\mathcal{P}_n$.
The codespace is defined as the simultaneous $+1$ eigenspace of all $S\in \mathcal{S}$.
For a stabilizer group generated by $n-k$ independent Pauli operators $S_1, \dots, S_{n-k}$, the dimension of the codespace is $2^k$, meaning it encodes $k$ qubits. 

When considering the action of a Pauli on a codeword, the phase of the operator is clearly physically irrelevant, so we will consider such a Pauli $e$ to be $\in\ppgroup{n}$.
By the definition of the stabilizer group, any $e\in\pstab$, where $\pstab:=\pi(\mathcal S)$, will act trivially on any codeword.
The information stored in the codespace is acted on by the logical operators of the code, the subgroup $\mathcal \plog := \pi (\mathcal{L}) \subset \ppgroup{n}$ where $\mathcal{L} := N(\mathcal S)/ \mathcal S$ is the normalizer of the stabilizer group in the Pauli group, modulo the stabilizers themselves. Indeed, $\plog$ consists of all Pauli operators not in $\pstab$ which commute with all its elements. 

Another important related group are the \emph{destabilizers} $\pdestab := \pi\left( N(\mathcal L)/ \mathcal S \right) \subset \ppgroup{n}$, those Pauli operators which commute with the logicals but which are not in the stabilizer group $\pstab$~\cite{aaronson_improved_2004}. 
Using a symplectic Gram-Schmidt procedure, any set of independent stabilizer generators can be extended to a symplectic basis 
in which  $z_1,\dots,z_k$ and $x_1,\dots,x_k$ generate $\plog$, $z_{k+1},\dots,z_n$ generate $\pstab$, and $x_{k+1},\dots,x_n$ generate $\pdestab$~\cite{wilde_logical_2009}. 
This symplectic basis extension of the stabilizer generators is sometimes called a \emph{tableau} of the code.
It should be noted that a tableau of a stabilizer code is not unique. 

As a consequence, given some tableau of a stabilizer code, any Pauli operator $q\in \ppgroup{n}$ can be uniquely decomposed in a stabilizer, destabilizer and logical part, i.e. $q=s\cdot d\cdot l$ where $s\in\pstab$, $d\in\pdestab$, and $l\in\plog$.
More concretely, we can write $q=x_1^{\lambda_1}\cdot \dots \cdot x_n^{\lambda_n} \cdot z_1^{\mu_1}\cdot \dots z_n^{\mu_n}$ where $\lambda_i=\langle q, z_i\rangle$ and $\mu_i=\langle q, x_i\rangle$.

The general error correction procedure for a stabilizer code can be described as follows.
First, the stabilizer generators $S_1,\dots,S_{n-k}$ are measured, each one yielding a result $m_j\in \mathbb F_2$. 
Because the stabilizers commute, the order of the measurement does not matter.
The entire collection of outcomes, $m\in \mathbb F_2^{n-k}$ is called the \emph{syndrome}.
The next step of the correction procedure is to decide on a suitable Pauli correction operation given the observed syndrome. 
This requires only classical computation, which we will call the \emph{decoder}.
Finally, the selected Pauli correction operation is applied to the data qubits.

\subsection{Optimal Pauli Noise Decoding}\label{sec:optimal_decoding}
The correction procedure succeeds if the product of the actual error $e\in\ppgroup{n}$ and the correction operation $e'\in\ppgroup{n}$ is a stabilizer: $e'e\in \pstab$.
The fact that successful correction need not determine the actual error is due to the degeneracy of quantum stabilizer codes. 

Decomposing $e=s\cdot d\cdot l$ and $e'=s'\cdot d'\cdot l'$, it is apparent that the stabilizer contribution is irrelevant. That is, $e''=se'$ is just as good a correction operation as $e'$ for any $s\in \pstab$.
Meanwhile, the syndrome $m$ specifies the destabilizer contribution as $d=x_{k+1}^{m_1}\cdot \dots \cdot x_{n}^{m_{n-k}}$ since the $s$ and $l$ contributions commute with $\pstab$. 
Therefore to return to the codespace (with trivial syndrome), it is necessary that $d'=d$. 

The remaining task of the decoder is to select an appropriate $l'$ given the syndrome $m$. 
Suppose that the Pauli errors occur according to some distribution $P_E$ over $\ppgroup{n}$, so that the probability of an error $e$ is $P_E(e)$. 
In view of the decomposition into stabilizer, destabilizer, and logical operator contributions, the distribution over errors may be thought of as a distribution over the random variables $S$, $D$, and $L$: $P_{S,D,L}(s,d,l):=P_E(s\cdot d\cdot l)$. 
The optimal decoding strategy then consists of selecting $l'$ to be the most likely logical operator given the observed syndrome $m$. 
This may be written as 
\begin{align}
l'=\argmax\limits_{l\in \plog} P_{L|D=d(m)}(l)\,,
\end{align}
where we have written the destabilizer $d$ as a function of $m$ since the latter determines the former.  
By Bayes' rule we have $P_{L|D=d(m)}(l)=P_{L,D}(l,d(m))/P_D(d(m))$, and therefore 
\begin{align}
\label{eq:optimal_decoder}
l'=\argmax\limits_{l\in \plog} \sum_{s\in \pstab} P_{S,D,L}(s,d(m),l)=\argmax\limits_{l\in \plog} \sum_{s\in \pstab} P_{E}(s\cdot d(m)\cdot l)\,.
\end{align}

The optimal decoder is only concerned with the most-likely error class, namely Pauli operators $s\cdot d\cdot l'$ for any $s\in \pstab$. 
It is not concerned with the most-likely error, given by $s'\cdot d\cdot l'$ for $(s',l')=\argmax_{s\in \pstab,l\in\plog}P_{E}(s\cdot m\cdot l)$.

\subsection{Circuit-level Noise}
When realizing error correction experiments in practice, one encounters the additional difficulty that the syndrome measurement itself is also affected by noise.
This significantly complicates the decoding process on one hand because the syndrome measurement outcome cannot be trusted (and thus the destabilizer part of the error cannot be determined with certainty) and on the other hand because the syndrome readout circuit itself might generate and spread noise throughout the data qubits.
The most common solution to address this issue for 2D topological codes is to repeat the syndrome measurement multiple times, which essentially increases the dimensionality of the decoding problem from 2D to 3D~\cite{dennis_topological_2002}.

While the stabilizer formalism provides a means to describe the relationship between error events in the circuit and observed measurement outcomes, it can be computationally advantageous to make use of the slightly different formalism of \emph{detector error models} (DEMs)\cite{gidney_stim_2021}.
In a DEM, noise in the circuit is modelled as a collection of \emph{error mechanisms}, where each ``mechanism'' is a  a single Pauli operator occurring somewhere in the circuit. Furthermore, these errors are modelled as occurring independently of each other. 
For example, depolarizing noise can be represented in the framework of DEMs, as it can be seen as the concatenation of  independent $X$, $Y$ and $Z$ error mechanisms, each occurring with identical probability:
\begin{equation}
    \mathcal{D}_{p} = \left((1-r_q)\cdot [\mathds{1}] + r_q\cdot [\sigma_X]\right) \circ \left((1-r_q)\cdot [\mathds{1}] + r_q\cdot [\sigma_Y]\right) \circ \left((1-r_q)\cdot [\mathds{1}] + r_q\cdot [\sigma_Z]\right)
\end{equation}
where $\mathcal{D}_{p}(\rho) := (1-p)\rho + p\frac{\id}{2}$ is the depolarizing channel of rate $p$, $r_q:=\frac{1-\sqrt{1-p}}{2}$, and $[U](\rho):=U\rho U^{\dagger}$ denotes the channel induced by the unitary $U$.
Note that not all single-qubit Pauli channels can be represented as three independent $X$, $Y$ and $Z$ error mechanisms.
Compared to the stabilizer formalism in which individual qubit Pauli errors are specified by an element of $\mathbb F_2^2$, e.g.\ $X^uZ^v$ for $(u,v)\in \mathbb F_2$, the DEM framework uses an overcomplete representation $X^uY^vZ^w$ for $(u,v,w)\in \mathbb F_2^3$. Thus $(u,v,w)$ and $(1-u,1-v,1-w)$ specify the same Pauli error. 

Every error mechanism in a DEM causes certain \emph{detectors} to flip, where detectors are parity checks of measurement outcomes that would be zero should no noise occurs in the circuit.
Often the detectors are given by the parity of two-subsequent measurements of the same stabilizer.
Typically, detectors should be chosen in such a way that an error mechanism only causes detectors to flip that are nearby in space and time.
Similarly to detectors, an error mechanism can also cause a \emph{logical observable} to flip.
Analogously to detectors, these are parity checks that are supposed to take a deterministic value if no noise is present in the circuit.
The main difference is that logical observables are not given as input to the decoding algorithm---instead the task of the decoder is to estimate their value.

Mathematically speaking, a DEM with $n$ error mechanisms and $m$ detectors is characterized by the triple $(H,p,\ell)$, with $H\in\mathbb{F}_2^{m\times n}$ a parity check matrix, $p\in \mathbb R^n$  a probability vector, and $\ell\in \mathbb F_2^n$ a logical error indicator vector. 
Here, we have assumed a DEM with only a single logical observable, but the formalism can be extended to multiple logical operators.
The entry $H_{i,j}\in \mathbb F_2$ is $1$ if and only if the error mechanism $j$ causes the detector $i$ to flip and zero otherwise, while the $i$th component $p_i$ of $p$ is the probability that the $i$th error mechanism occurs, and the $i$th component $\ell_i$ of $\ell$ is 1 when the $i$th error mechanism causes the logical observable to flip and 0 otherwise.

It should be noted that multiple error mechanisms which cause the identical detectors to flip and have the same logical effect can be combined into one single error mechanism with combined probability.
This optimization is always done in practice, as it can considerably reduce the number of error mechanisms. Indeed, this is one of the main advantages of the DEM formulation. 

The optimal decoder for a detector error model is formulated as follows:
Assume that the measurement outcome of the detectors is given by $m\in\mathbb{F}_2^m$.
The joint probability of the detector outcome $m$ occurring and the logical error being $L\in \mathbb F_2$ is given by
\begin{equation}
    p_{m,L} = \sum\limits_{\substack{x\in\mathbb{F}_2^n \\ Hx=m \text{ and } \\ \ell\cdot x=L}} P_X[x]
\end{equation}
where
\begin{equation}
    P_X[x] := (1-p_1)^{1-x_1}p_1^{x_1}\cdots (1-p_n)^{1-x_n}p_n^{x_n} \, .
\end{equation}
The optimal logical correction $L^*$ is then given by $L^*(m) = \argmax\limits_{L\in\mathbb{F}_2} p_{m,L}$ due to a similar reasoning as in~\Cref{sec:optimal_decoding}.

\subsection{Tensor Networks}
Here, we give a brief overview; for more details see e.g.~\cite{bridgeman_hand-waving_2017}.
Fundamentally, tensor networks are a graphical representation of algebraic expressions which are sums of products. 
This graphical representation can be useful in designing algorithms for exact or approximate evaluation of the intended expression. 
Consider an expression of the form $g(x_1,x_2)=\sum_{y_1,y_2,y_3} f_1(x_1,y_1,y_3)f_2(x_2,y_2,y_3)f_3(y_1,y_2)$. 
We associate a vertex or node to each factor and to each variable external to the summation. 
Edges connect the appropriate nodes; there is one edge for each of the internal and external variables. 
The result is the following.
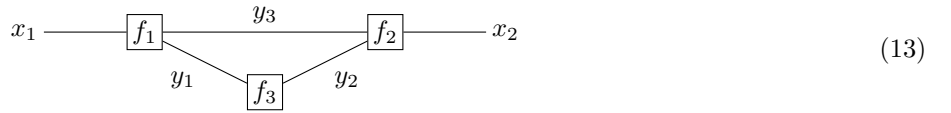
\begin{equation}
\input{figures/simpleTN}
\end{equation}
Our convention is edges terminating on a variable name fix the value of that edge, so the tensor network represents $g(x_1,x_2)$ for given values of $x_1$ and $x_2$. 
On the other hand, a variable name next to an edge is just a label. 
In this way we can use the tensor network to represent the entire function $g$ by using half-edges for free variables: 
\begin{equation}
\input{figures/simpleTN2}
\end{equation}

The example $g$ is not the general case of a sum-of-products, as no variable appears in more than two factors. 
However, any general sum of products can be brought into this particular form by making use of additional internal variables and indicator functions to enforce equality. 
For instance, consider
\begin{equation}
g'(x_1,x_2)=\sum_{y_1,y_2,y_3} f_1(x_1,y_1,y_3)f_2(x_2,y_2,y_3)f_3(y_1,y_2,y_3),   
\end{equation}
where $y_3$ now appears in each factor. 
But this is just
\begin{equation}
g'(x_1,x_2)=\sum_{y_1,y_2,y_3,z_1,z_2} f_1(x_1,y_1,z_1)f_2(x_2,y_2,z_2)\,f_3(y_1,y_2,y_3)\delta(y_3,z_1,z_2)
\end{equation}
for $\delta$ the function which is 1 when all its arguments are equal and zero otherwise.  
Now we apply the above construction. 
Denoting the $\delta$ factor by $=$, the following tensor network represents the function $g'$:
\begin{equation}
\input{figures/simpleTN3}
\end{equation}

\section{Optimal Decoding as Tensor Network Contraction}
\label{sec:optimaldecoding}
In this section we describe two formulations of the quantity $\sum_{s\in \pstab}P_{E}(s\cdot d(m)\cdot l)$ as tensor network contraction. 
By~\Cref{eq:optimal_decoder}, this is sufficient to realize the optimal decoder; the error class probability can be calculated for each value of $l$ and fixed $m$ via contraction. 
We call the two formulations the \emph{detector picture} and the \emph{\generator{} picture}. 

Specifically, we will consider the case of independent single-qubit noise of the form
\begin{equation}
P_E(q)=P_1(q^{(1)})\cdot P_2(q^{(2)})\cdots P_n(q^{(n)}),    
\end{equation}
where the $P_i$ for $i\in \{1,\dots n\}$ are distributions over $\{\mathds{1},\sigma_X,\sigma_Y,\sigma_Z\}$. 
This will ensure that the tensor network inherits the topology of the stabilizer code, i.e.\ a 2D topological code results in a tensor network that is local in 2 dimensions and a 3D topological code results in a tensor network that is local in 3 dimensions. 
More generally, it would be sufficient to assume that the noise model is expressible as a Markov random field~\cite{chubb_statistical_2021}, but we will not pursue this issue further here. 

Exact contraction of the resulting tensor networks will generally be infeasible, and therefore approximate contraction procedures are required in practice. 
These are further discussed in~\Cref{sec:approximate_contraction}.

Furthermore, we also discuss how the \generator{} and detector picture simplify for the special case of CSS codes under purely $X$-type and $Z$-type Pauli noise.
Finally, we also briefly discuss how to extend the formalism to circuit-level noise, which is more relevant for experimental implementations of quantum error correction.

Both tensor network formulations begin with the expression
\begin{equation}
\sum_{s\in \mathcal \pstab}P_{E}(s\cdot d(m)\cdot l) = \sum_{q\in \ppgroup{n}} P_E(q)\,\Pi_{m,l}(q)
\end{equation}
where $\Pi_{m,l}(q)$ is an indicator function enforcing that $q$ should have syndrome $m$ and logical component $l$. 
This is the basic sum-of-products expression that leads to a tensor network representation. 
It will be more convenient to represent each Pauli operator as an element of $\mathbb F_2^{2n}$ using the map $w$ instead of working over $\ppgroup{n}$. 
Abusing notation somewhat to write $P_E(y)$ with $y\in \mathbb F_2^{2n}$ for $P_E(q)$ with $y=w(q)$ and similarly for $\Pi_{m,l}$, we have
\begin{equation}
\label{eq:tnstart}
\sum_{s\in \pstab}P_{E}(s\cdot d(m)\cdot l) = \sum_{y\in \mathbb F_2^{2n}} P_E(y)\,\Pi_{m,l}(y)\,.
\end{equation}
Two different ways of expressing the indicator function give rise to the two tensor network formulations. 
Both make use of the parity function $\mathbb P$, which is zero unless the sum (modulo 2) of its arguments is itself zero.

\subsection{Detector Picture}
\subsubsection{Initial formulation}
In the detector picture we start by using the parity function to enforce the particular syndrome and logical contribution. 
Working directly in the $\mathbb F_2^{2n}$ representation, the syndrome constraint is simply $y\cdot \bar w(s_i)= m_i$ for $i=1,\dots,n-k$, where we have defined $\bar w(q)=Jw(q)$ for any Pauli operator $q$ and $s_1,\dots,s_{n-k}$ denote the stabilizer generators modulo phase.
These constraints are captured by the indicator function $\prod_{i=1}^{n-k}\mathbb P(y\cdot \bar w(s_i), m_i)$. 
The constraint for the logical operator is similar. 
Writing $l$ in terms of the generators $x_j$ and $z_j$ as $l=x_1^{a_1}\cdot x_{2}^{a_2}\cdots x_n^{a_n}\cdot z_1^{b_1}\cdot z_2^{b_2}\cdots z_n^{b_n}$ with $a,b\in \mathbb F_2^{k}$, we then have $\prod_{j=1}^{k}\mathbb P(y\cdot \bar w(x_j), b_j)\mathbb P(y\cdot \bar w(z_j),a_j)$.
Note that commutation or anticommutation with $x_j$ determines the $z_j$ contribution to the Pauli operator, and vice versa. 
Thus we have 
\begin{equation}
\label{eq:detectorind}
\Pi_{m,l}(q)=\prod_{i=1}^{n-k}\mathbb P(y\cdot \bar w(s_i), m_i)\prod_{j=1}^{k}\, \mathbb P(y\cdot \bar w(x_j), b_j)\,\mathbb P(y\cdot \bar w(z_j),a_j)\,.
\end{equation}

Using this expression in~\Cref{eq:tnstart} gives a summation over product terms.
For convenience we write the summed variable $y=(u,v)$ where $u\in\mathbb F_2^n$ is the $X$ component and $v\in\mathbb F_2^n$ is the $Z$ component of the error.
Then for each qubit the TN includes a tensor node for the probability of error on that qubit.
We label the node for the $i$th qubit $P_i$ and it has two edges, corresponding to $u_i$ and $v_i$, specifying the $X$ and $Z$ components of the Pauli error on that qubit, respectively. 

For the parity factors, first observe that since $y\cdot \bar w(s_i)$ is just the sum of the components of $y=(u,v)$ for which $\bar w(s_i)$ is not zero, we may simply take these components as individual arguments to $\mathbb P$. 
Thus, each parity function is associated to a tensor node with edges given by the appropriate components of $u$ and $v$, as well as a bit coming from the constraints, be it $m_i$, $a_i$, or $b_i$. 
Tensor nodes associated to parity functions are called check nodes and labelled by `$+$'.

Since any given $u_i$ or $v_i$ may participate in multiple parity constraints, to connect the output of a probability tensor $P_j$ we make use of an equality tensor to copy the values of $u_i$ and $v_i$. 
An equality node, labelled by `$=$', simply represents the function which is 1 if all of its arguments agree, and zero otherwise.

\begin{figure}
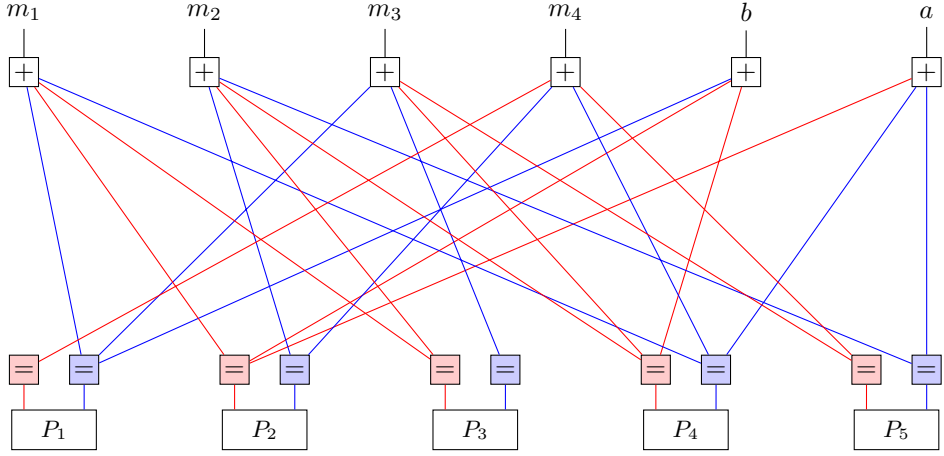

    \centering
    \include{figures/tn_detector_col}
    \caption{Detector picture tensor network of the $5$-qubit code. In order to improve readability, the lines in the network are coloured red and blue to indicate connections to the $X$ and $Z$ tensors.}
    \label{fig:5qbcode_detector_tn}
\end{figure}

\Cref{fig:5qbcode_detector_tn} depicts the detector tensor network associated to the decoding problem of the 5-qubit code. 
The stabilizer generators are taken to be $s_1=XZZXI$, $s_2=IXZZX$, $s_3=XIXZZ$, $s_4=ZXIXZ$ and logical generators $x_1=XZIZI$, $z_1=IZIXX$.
Edges terminating in variables are understood to take only the value of the indicated variable, i.e.\ an edge terminating in `$a$' takes only the value $a$. 
The equality nodes are colored red and blue to distinguish $u_i$, or $X$ contributions, in red from $v_i$, $Z$ contributions, in blue.

To summarize, the detector picture tensor network is obtained as follows.
\begin{framed}
The tensor network is specified by the qubit probability distributions $\{P_j\}_{j=1}^n$, the syndrome $m\in \mathbb F_2^{n-k}$, the associated stabilizer generators $\{s_i\in \pstab\}_{i=1}^{n-k}$, the logical generators $\{x_j,z_j\in \plog\}_{j=1}^k$, and a logical operator $l\in \plog$ with $l=x_1^{a_1}\cdot x_{2}^{a_2}\cdots x_n^{a_n}\cdot z_1^{b_1}\cdot z_2^{b_2}\cdots z_n^{b_n}$ with $a,b\in \mathbb F_2^{k}$.
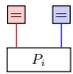
\begin{enumerate}
    \item The $i$th qubit is represented by 
        \scalebox{0.6}{
        \begin{tikzpicture}[square/.style={regular polygon,regular polygon sides=4,inner sep=0,minimum height=15}]
            \node[square,draw,fill=red!20] (x) at (0,0) {\eq};
            \node[square,draw,fill=blue!20] (z) at (1,0) {\eq};
            \node[minimum height=15,minimum width=40,draw,fill=white] (p) at (0.5,-1) {\small $P_i$};
            \draw[red] (x) -- (p.north-|x);
            \draw[blue] (z) -- (p.north-|z);
        \end{tikzpicture}
        }
    with the red and blue equality nodes corresponding to the $X$ and $Z$ components respectively.
    \item The $j$th stabilizer generator is represented by a check node with one edge connected to the syndrome value $m_j$.  
    \item The $j$th logical generator $x_j$ ($z_j$) is represented by a check node with one edge connected to $b_j$ ($a_j$).
    \item A check node is connected to an equality node (of $X$ or $Z$ type) if and only if the corresponding qubit is involved in the corresponding stabilizer or logical generator. For instance, the $j$th stabilizer check is connected to $u_i$ if $\bar w(s_j)_i=1$ and $v_i$ if $\bar w(s_j)_{i+n}=1$. (Note the use of $\bar w$ here.)
\end{enumerate}
\end{framed}

\subsubsection{Removing high-weight logical parity nodes}\label{sec:hadamard_trick}

In the setting of topological codes, the check nodes corresponding to stabilizer generators are local and of constant degree. However, the check nodes associated to the logical generators, i.e.\ the factors $\mathbb P(y\cdot \bar w(x_j), b_j)$ and $\mathbb P(y\cdot \bar w(z_j),a_j)$ typically have neither of these properties. 
Their connectivity can cause difficulties in the approximate tensor contraction methods described in~\Cref{sec:approximate_contraction}. 
These difficulties can however be circumvented by using the Walsh-Hadamard transform.

Recall that the decoder will need to evaluate $2^{2k}$ tensor networks, one for each value of $a,b\in \mathbb F_2^{k}$. 
That is to say, we are ultimately interested in the tensor network like that of~\Cref{fig:5qbcode_detector_tn} but where the logical check nodes have open edges instead of these edges terminating on fixed values $a$ and $b$. 
The network with open edges represents the collection of all the networks with fixed edges. 
Using the \emph{Hadamard tensor} we can instead determine all the requisite error class probabilities by evaluating $2^{2k}$ simpler tensor networks and subsequently performing a Walsh-Hadamard transform of the resulting data. 

The components of the degree-two Hadamard tensor on binary-valued edges are determined by the $2\times 2$ Hadamard matrix. 
Observe that contracting a separate Hadamard tensor to all the edges of an equality tensor results in a tensor proportional to the parity tensor. 
Algebraically, say for the degree-three case, this is the statement that
\begin{equation}
\sum_{x',y',z'\in \mathbb F_2}H_{xx'}H_{yy'}H_{zz'}\,\delta_{x'y'}\,\delta_{y'z'}\,\,\propto\,\, \mathbb P(x,y,z)\,.
\end{equation}
In tensor network notation this is just 
\begin{equation}
\input{figures/Hadamard}
\end{equation}

This equivalence allows us to replace the $2k$ parity check nodes associated with the logical generators with equality nodes and degree-two Hadamard nodes. 
For instance, if there are two logical parity check nodes, the general situation is given by 
\begin{align}
\input{figures/WalshHadamard}
\end{align}
where the cloud represents the remainder of the network. 
Instead of determining the values of the total tensor network, one may instead evaluate network inside the dashed box.
In this example, there are four such networks to be evaluated, and in general for $2k$ parity nodes there will be $2^{2k}$ relevant networks. 
Regarding this contraction data as a vector of $2^{2k}$ entries, the final tensor network contraction with the ``outbound'' Hadamard nodes is just the Walsh-Hadamard transform of the data. 

This approach fixes the problem of the non-local nodes, as the remaining high-weight equality node in the dashed network can be replaced with equality nodes on each involved site separately, given that the value of the external leg is fixed.

\subsection{Generator Picture}
The generator picture takes a different approach to handling the dependence on the logical operator $l$. 
Returning to~\Cref{eq:tnstart}, the required terms in the sum correspond to summing over all the Pauli operators in the set \mbox{$F_{m,l}=\{s\cdot x_{k+1}^{m_1}\cdots x_{n}^{m_{n-k}}\cdot l:s\in \pstab\}$} (a coset of the stabilizer group). 
Equivalently, for any $r_{m,l}$ in $F_{m,l}$, i.e.\ any Pauli operator with syndrome $m$ and logical component $l$, $F_{m,l}$ can equally-well be specified by \mbox{$\{sr_{m,l}:s\in \pstab\}$}.
In terms of the $\mathbb F_2^{2n}$ representation, $F_{m,l}$ is specified by \mbox{$\{w(r_{m,l})+\sum_{i=1}^{n-k}\lambda_i w(s_i):\lambda\in \mathbb F_2^{n-k}\}$} where $s_1,\dots,s_{n-k}$ denote the stabilizer generators. 
The indicator function $\Pi_{m,l}(y)$ enforces that $y$ is an element of this set. 
Therefore, in terms of the parity function $\mathbb P$, it holds that 
\begin{equation}
\Pi_{m,l}(y) 
=\sum_{\lambda\in \mathbb F_2^{n-k}}\mathbb P(y,w(r_{m,l}),\lambda_1 w(s_1),\dots,\lambda_{n-k}w(s_{n-k}))\,.
\end{equation}
Therefore, we now have the expression
\begin{equation}
\label{eq:startgen}
\sum_{y\in \mathbb F_2^{2n}} P_E(y)\,\Pi_{m,l}(y) = \sum_{y\in \mathbb F_2^{2n}} \sum_{\lambda\in \mathbb F_2^{n-k}}P_E(y) \, \mathbb P(y,w(r_{m,l}),\lambda_1 w(s_1),\dots,\lambda_{n-k}w(s_{n-k}))\,,
\end{equation}
which is also in sum-of-products form, and hence amenable to tensor network representation.
Here, we are using a slightly different parity function than in the previous section; in~\Cref{eq:startgen} the arguments are  elements of $\mathbb F_2^{2n}$, as opposed to single bits.
However, $\mathbb P(y,y')$ for $y=(u,v)$ and $y'=(u',v')$ both in $\mathbb F_2^{2n}$ can be decomposed into $\prod_{j=1}^n \mathbb P(u_j,u'_j)\mathbb P(v_j,v'_j)$. 

The resulting algebraic expression for $\sum_{y\in \mathbb F_2^{2n}} P_E(y)\,\Pi_{m,l}(y)$ is rather unwieldy, so let us  proceed directly to the tensor network description. 
In contrast to the detector picture, now there is a parity check for each $u_j$ and for each $v_j$ with edges heading to three different places. 
The $u_j$ and $v_j$ checks are each connected to the associated probability tensor $P_j$ (again assuming the distribution $P_E$ is independent). 
Another set of edges connects the $u_j$ and $v_j$ checks to the appropriate components of $w(r_{m,l})$; defining $w(r_{m,l})=(r^x,r^z)$ with $r^x$ and $r^z$ elements of $\mathbb F_2^n$, the $u_j$ node is connected to the value $r^x_j$ and the $v_j$ node to the value $r^z_j$.
Finally, since the $\lambda_j$ participate in many parity checks, there is an equality node for each $\lambda_j$, connected to those $u_i$ and $v_i$ for which $w(s_j)_i=1$ and $w(s_j)_{i+n}=1$, respectively.

\begin{figure}[h]
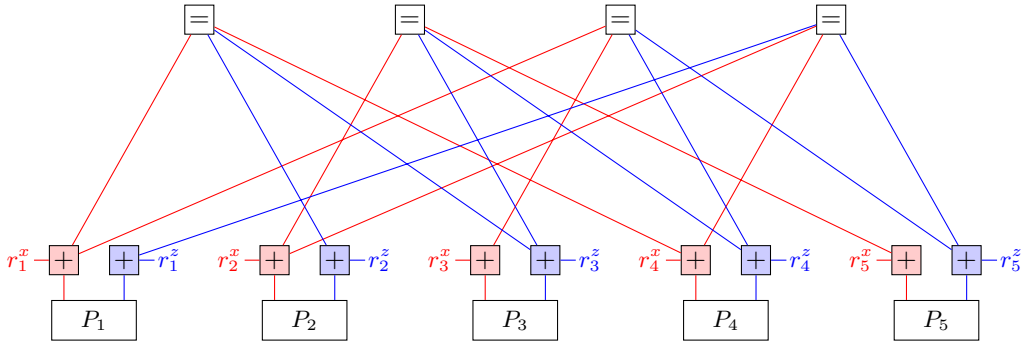

    \centering
    \include{figures/tn_gauge_col}
    \caption{Generator picture tensor network of the $5$-qubit code. As before, the lines in the network are coloured to improve readability.}
    \label{fig:5qbcode_generator_tn}
\end{figure}

\Cref{fig:5qbcode_generator_tn} depicts the resulting tensor network for the 5-qubit code, where we have chosen $r_{m,l}$ using only the logical operators and the destabilizer as suggested above. 
Observe that the values of $r_{m,l}$ could be locally incorporated into the check node tensors.

To summarize, the \generator{} picture tensor network is obtained as follows. 
\begin{framed}
The tensor network is specified by the qubit probability distributions $\{P_j\}_{j=1}^n$, a representative Pauli operator $r$ satisfying the syndrome $m\in \mathbb F_2^{n-k}$ and logical contribution $l$, and the stabilizer generators $\{s_i\in \pstab\}_{i=1}^{n-k}$.
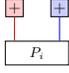
\begin{enumerate}
    \item The $i$th qubit is represented by 
    \scalebox{0.6}{
        \begin{tikzpicture}[square/.style={regular polygon,regular polygon sides=4,inner sep=0,minimum height=15},scale=1.0]
            \node[square,draw,fill=red!20] (x) at (0,0) {$+$};
            \node[square,draw,fill=blue!20] (z) at (1,0) {$+$};
            \node[minimum height=15,minimum width=40,draw,fill=white] (p) at (0.5,-1) {\small $P_i$};
            \draw[red] (x) -- (p.north-|x);
            \draw[blue] (z) -- (p.north-|z);
        \end{tikzpicture}
    } with the red and blue parity nodes corresponding to the $X$ and $Z$ components respectively.
    \item Parity node $u_i$ is connected to the value $w(r)_i$ and parity node $v_j$ to $w(r)_{i+n}$. 
    \item Each stabilizer generator $s_j$ is represented by an equality node.
    \item Equality node $s_j$ is connected to parity node $u_i$ when $w(s_j)_i=1$ and to parity node $v_j$ when $w(s_j)_{i+n}=1$. 
\end{enumerate}
\end{framed}

\subsection{Bit- and phase-flip Noise on CSS Codes}\label{sec:css_special_case}
Consider the special case where the considered code is CSS, i.e.\ all stabilizer generators are either purely $X$-type or $Z$-type.
We will assume that the logical generators are chosen such that they are also all $X$-type or $Z$-type.
Furthermore, assume that the noise model is given by independent $X$ and $Z$ errors on every qubit
\begin{aligns}
    P_i(\id) &= (1-p^x_i)(1-p^z_i) \\
    P_i(\sigma_X) &= p^x_i(1-p^z_i) \\
    P_i(\sigma_Z) &= (1-p^x_i)p^z_i \\
    P_i(\sigma_Y) &= p^x_ip^z_i \,
\end{aligns}
for some $p_i^x,p_i^z\in [0,1]$.
In terms of the tensor representation of the distribution of noise on a single qubit, this structure means that we can separate
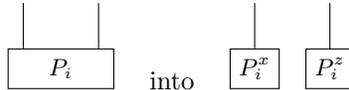
\begin{center}
\begin{tikzpicture}
    \node (x) at (0,0) {};
    \node (z) at (1,0) {};
    \node[minimum height=15,minimum width=40,draw,fill=white] (p) at (0.5,-1) {\small $P_i$};
    \draw (x) -- (p.north-|x);
    \draw (z) -- (p.north-|z);
\end{tikzpicture}
\quad into \quad 
\begin{tikzpicture}
    \node (x) at (0,0) {};
    \node (z) at (1,0) {};
    \node[minimum height=15,minimum width=15,draw,fill=white] (x2) at (0,-1) {\small $P_i^x$};
    \node[minimum height=15,minimum width=15,draw,fill=white] (z2) at (1,-1) {\small $P_i^z$};
    \draw (x) -- (x2.north-|x);
    \draw (z) -- (z2.north-|z);
\end{tikzpicture}
\end{center}
where $P_i^x$ and $P_i^z$ represent the vectors $(1-p_i^x,p_i^x)$ and $(1-p_i^z,p_i^z)$.

Under these assumptions the tensor network (both in the detector and generator pictures) separates into two disjoint sub-networks.
One network is responsible for the decoding decision for $Z$-type logical operators (i.e.\ the problem of decoding bit-flip errors) and the other is responsible for the decoding decision for $X$-type logicals.
To see this, notice that in the detector picture, the parity checks only act on one of the two sub-networks.
Similarly, the logical component of the representative chosen in the generator picture can be separated into $X$-type and $Z$-type logicals, which again only impact the respective sub-network.

In the case when either all $p_i^x$ are zero or all $p_i^z$ are zero, one of the two sub-networks trivially contracts to $1$ and can be omitted, meaning that only the other sub-network remains.
More concretely, only one of the two stabilizer types (either $X$ or $Z$ type) remains represented as nodes in the tensor network.
Interestingly, depending on whether one is in the detector or generator picture, the remaining type is different.
For instance, when one considers pure bit-flip noise on a CSS code, then the detector picture involves the $Z$-type stabilizers, whereas the generator picture involves the $X$-type stabilizers.
If we incorporate the remaining probability tensors into the neighboring $=$ or $+$ nodes, the remaining graph is a bipartite graph with one subgraph containing $=$ nodes and the other $+$ nodes.
In the detector picture under bit-flip noise, the connectivity of the bipartite graph is described by the $Z$-type parity check matrix matrix $H_Z$ of the CSS codes, whereas in the generator picture under bit-flip noise the connectivity is described by a dual matrix $G_Z$ which fulfills $H_Z\cdot G_Z^T=0$.
This formalizes the notion of the generator picture being the dual of the detector picture.

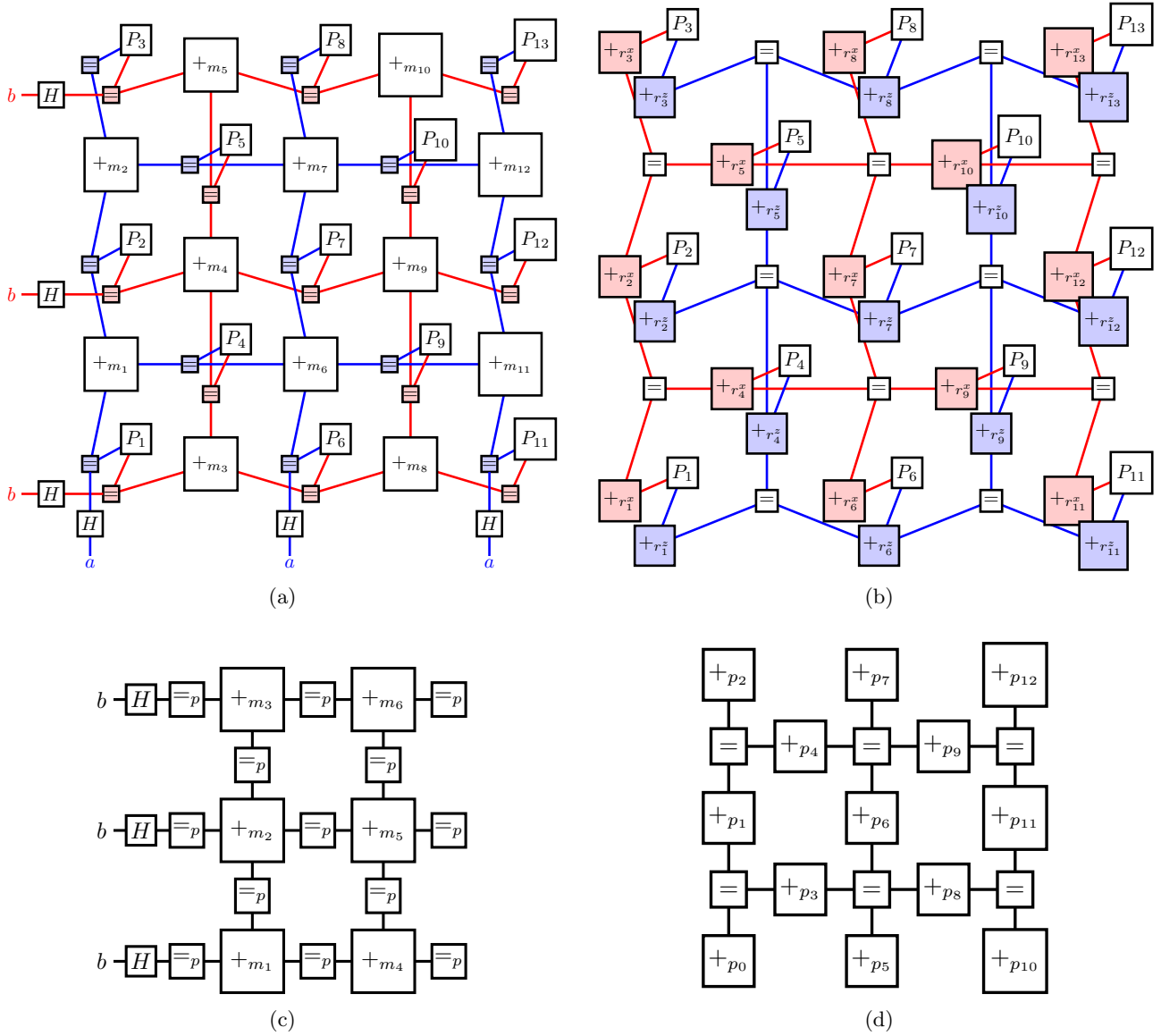
\begin{figure}
     \begin{subfigure}[b]{0.50\textwidth}
          \centering
          \resizebox{0.95\linewidth}{!}{\input{figures/sc2d_detector_col}}  
          \caption{}
          \label{fig:sc2d_depol_detector}
     \end{subfigure}
     \begin{subfigure}[b]{0.50\textwidth}
          \centering
          \resizebox{0.95\linewidth}{!}{\input{figures/sc2d_generator_col}}  
          \caption{}
          \label{fig:sc2d_depol_generator}
     \end{subfigure}
     ~\\
     \begin{subfigure}[b]{0.50\textwidth}
          \centering
          \resizebox{0.65\linewidth}{!}{\input{figures/sc2d_bitflip_detector}}  
          \caption{}
          \label{fig:sc2d_bitflip_detector}
     \end{subfigure}
     \begin{subfigure}[b]{0.50\textwidth}
          \centering
          \resizebox{0.6\linewidth}{!}{\input{figures/sc2d_bitflip_generator}}  
          \caption{}
          \label{fig:sc2d_bitflip_generator}
     \end{subfigure}
     \caption{Detector (a) and generator (b) tensor networks for d=3 unrotated surface code for an arbitrary i.i.d. single-qubit Pauli noise model with marginals $P_i$. Detector (c) and generator (d) tensor networks for d=3 unrotated surface code under bit-flip noise of strength $p$. Here $=_q$ is the tensor that takes value $(1-q)$ if all incoming legs have value $1$, value $q$ if all incoming legs have value $0$, and value $0$ otherwise. Also, $+_q$ is the tensor that takes value $(1-q)$ if the incoming legs sum up to 0 (mod 2) and $q$ otherwise. In (d) the $p_i$ are defined as $p$ if the $i$-th entry $r_i^x$ of the representative bit-flip error is $0$, and $1-p$ otherwise. The detector tensor networks in (a) and (c) implement the Walsh-Hadamard transform discussed in~\Cref{sec:hadamard_trick}.}
 \end{figure}

As an illustrative example,~\Cref{fig:sc2d_bitflip_detector,fig:sc2d_bitflip_generator} depict the detector and generator tensor networks for the $d=3$ unrotated 2D surface code under bit-flip noise.
Only half of the plaquettes are involved in the diagram, in contrary to the detector and generator networks for general single-qubit i.i.d.\ noise,  which are depicted in~\Cref{fig:sc2d_depol_detector,fig:sc2d_depol_generator}.

\subsection{Detector Error Models for Circuit-level Noise}\label{sec:dem}
Mathematically speaking, the optimal decoding problem for a DEM is equivalent to the optimal decoding of one sector of a CSS code, with the corresponding parity check matrix given by the DEM.
The error mechanisms play the role of the qubits and the detectors play the role of the stabilizers generators.
Thus, the detector and generator tensor network constructions are directly applicable.
The resulting tensor network will be local under the assumption that all error mechanisms in the model are local, i.e. only trigger nearby detectors.
In circuit-level noise, this is usually achieved by defining detectors to be the parity of two subsequent measurements of the same stabilizer generator.

\section{Decoding by Approximate Contraction}\label{sec:approximate_contraction}
\label{sec:approxcontract}
The previous section laid out in detail how the optimal decoding process can be expressed in terms of a sequence of contractions of tensor networks that exhibit a topology respecting the locality of the underlying code.
The exact contraction of these tensor networks for larger codes is generally not feasible, so approximate contraction schemes must be used for a practical decoder.
While generally NP-hard, the approximate contraction of planar two-dimensional tensor network is theoretically well understood and tends to behave very well when used in practice.
In the context of decoding of 2D Pauli codes, previous work by one of us demonstrated a powerful approximate contraction scheme that proved extremely successful in decoding a wide variety of two-dimensional stabilizer codes and effectively achieved the optimal threshold in all cases~\cite{chubb_general_2021}.

Generally, three-dimensional approximate contraction is significantly less well-behaved than its two-dimensional counterpart.
In essence, the problem lies in the impossibility to define a canonical gauge for a tensor network that is not a tree~\cite[\S5.2]{orus_practical_2014}.
Still, there are some techniques that go beyond na\"ive time-evolving block decimation (TEBD) to deal with three-dimensional networks, with a certain amount of success~\cite{jahromi_universal_2019,corboz_simulation_2010,jiang_accurate_2008,tindall_gauging_2023,murg_variational_2007,lubasch_algorithms_2014,corboz_simulation_2010,phien_infinite_2015,jordan_classical_2008}.
These techniques are typically studied in the context of real or imaginary time evolution of two-dimensional condensed-matter systems, but they can also be applied in the setting of quantum error correction.
They incur different trade-offs between accuracy and speed.

The technique we present works as follows: We assume that the 3D tensor network can be written as a stack of layers where each layer has the identical topology and connections between layers may only occur between equivalent sites.
While the detector/generator tensor networks of most codes do not directly fulfill this assumption, we can slightly adapt the networks to make it hold (see~\Cref{sec:results} for details).
We sweep a 2D tensor network from bottom to top in a TEBD-style fashion, contracting in one layer after the other while continually truncating the 2D tensor network bonds.
For example, if the 3D tensor network is a cubic lattice, then we would sweep a PEPS from one side of the cube to the opposing side.
The remaining two-dimension network is then contracted by sweeping an MPS across it (in case it is not a square lattice, the sweep-line-based contraction algorithm from~\cite{chubb_general_2021} can be used).

\begin{figure}
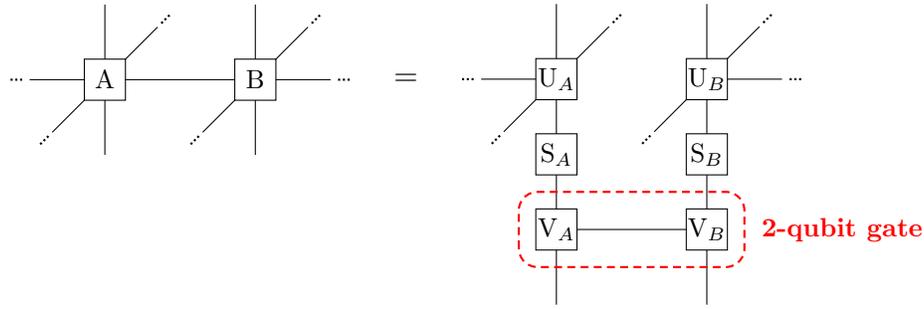

    \centering
    \include{figures/gate_splitting}
    \caption{By iteratively repeating the depicted procedure, a layer from the 3D tensor network can be split into a sequence of two-qubit gates. The decomposition of $A$ into $U_A,S_A$ and $V_A$ (respectively $B$ into $U_B,S_B$ and $V_B$) is done via the singular value decomposition. The contraction of $V_A$ and $V_B$ form the 2-qubit gate, while the contraction of $U_A$ and $S_A$, respectively $U_B$ and $S_B$ replace $A$ and $B$ in the layer. For practical applications, the number of nonzero eigenvalues in $S_A$ and $S_B$ are truncated to a maximal value to avoid encountering tensors that are too large.}
    \label{fig:splitting}
\end{figure}

A layer is contracted into the 2D tensor network by decomposing it into a sequence of two-qubit gates which can then be contracted one at a time.
For some tensor networks (when the involved tensors are weighted delta or check nodes for instance) this can be done in a natural way---this is the case for the 3D surface code under bit-flip or phase-flip noise, as will be discussed in the next section.
If no structure is known for doing the decomposition, one can instead use the singular value decomposition to separate a two-qubit gate from the layer, as depicted in figure~\Cref{fig:splitting}.

For the purpose of truncating the bond dimensions of the intermediate 2D tensor network state, we make use of the so-called \emph{simple update} technique~\cite{jahromi_universal_2019,corboz_simulation_2010,jiang_accurate_2008}.
The optimal truncation of a bond in a tensor network generally requires knowledge about the rest of the network.
More concretely, one needs to know the environment of the two nodes involved in the bond, i.e., the result of contracting the full network surrounding the nodes.
In the case of a tree tensor network, computing the environment can be avoided by bringing the network into its canonical gauge.
Unfortunately, such a procedure is not possible for loopy tensor networks, and computing the environment generally entails an exponential overhead.
Thus, truncation techniques, such as simple update, attempt to approximate the environment in some manner.
Simple update addresses this issue by associating to each leg of the tensor network a diagonal square matrix with size equal to the bond dimension.
For a given node in the network, the adjacent diagonal matrices provide a rank-1 approximation of its environment.
Under the assumption that this approximation is correct, the optimal truncation can be performed using a truncated singular value decomposition.
When a two-qubit gate is contracted into the PEPS and the corresponding bond then truncated, the involved environment tensors are correspondingly updated to reflect the change in the network.
For a step-by-step exposition of the simple update procedure we refer the reader to excellent explanations in the literature, such as~\cite[Appendix A]{patra_efficient_2023}.

As its name implies, simple update is a conceptually and computationally simple technique. It was recently shown to be sufficient for fast and highly accurate simulations of 2D quantum systems with more than a thousand qubits~\cite{patra_efficient_2023}, though it is less accurate compared to more computationally expensive approaches.

\section{Results}\label{sec:results}
\subsection{Point Sector of 3D Surface Code}
\begin{figure}
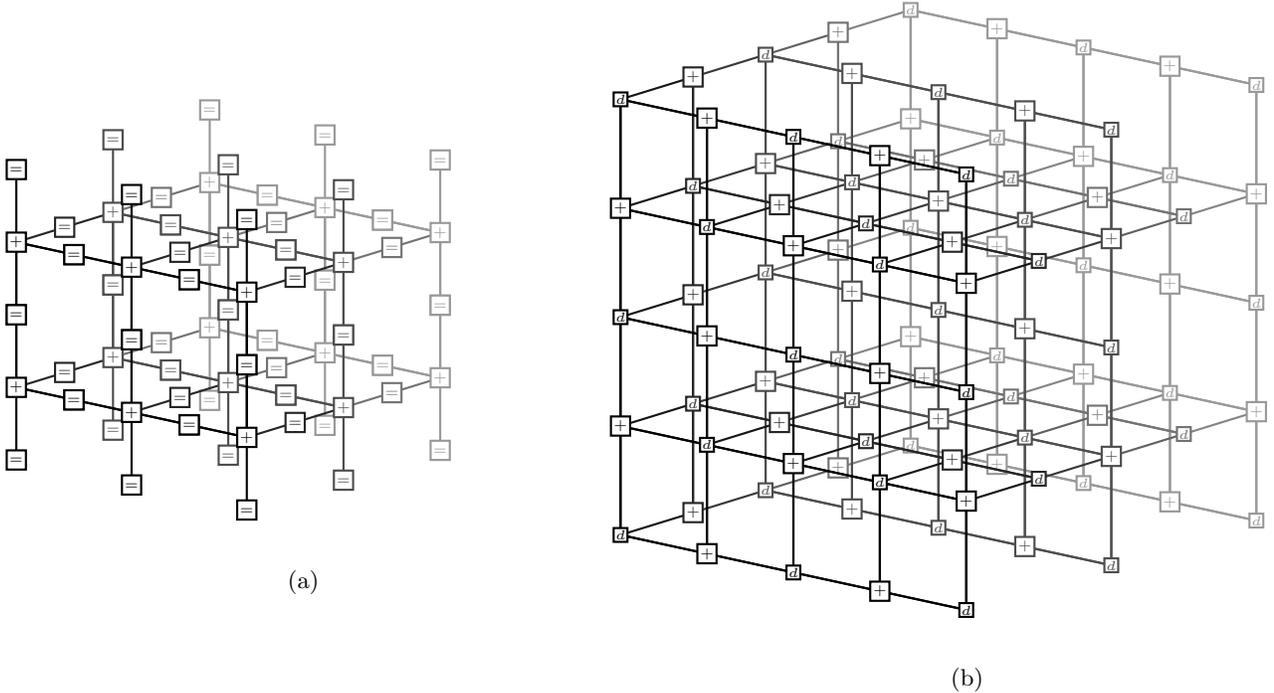

    \centering
    \begin{subfigure}[c]{0.45\textwidth}
        \tiny
        \include{figures/tn_pointsec}
        \normalsize
        \caption{}
        \label{fig:tn_pointsec}
    \end{subfigure}
    \begin{subfigure}[c]{0.54\textwidth}
        \tiny
        \include{figures/tn_depol}
        \normalsize
        \caption{}
        \label{fig:tn_depol}
    \end{subfigure}
    \caption{(a) Detector tensor network for the point sector of the 3D surface code of distance $d=3$.
    The $=$ and $+$ tensors are weighted (not displayed in figure), since the probability tensors, Hadamard tensors and syndrome values are incorporated in them.
    (b) Detector picture tensor network for the distance-$3$ 3D surface code subject to  depolarizing noise. The $d$ tensors represent the contraction of the single-qubit marginals $P_i$, the two connected $=$ tensors as well as the Hadamard gate for the Hadamard trick (where applicable). The syndrome values are incorporated in the $+$ nodes.}
    \label{TODO}
\end{figure}

Following the construction described in~\Cref{sec:css_special_case}, one can straightforwardly produce a detector and generator tensor network.
Consider the detector tensor network depicted in~\Cref{fig:tn_pointsec}: it is almost of a cubic tensor network form, except for some additional tensors on the bonds between the cubic lattice sites.
In contrary, the generator picture is less favorable and cannot be embedded as easily in a cubic lattice of the same size.
For this reason, we make use of the detector picture here.

\begin{figure}
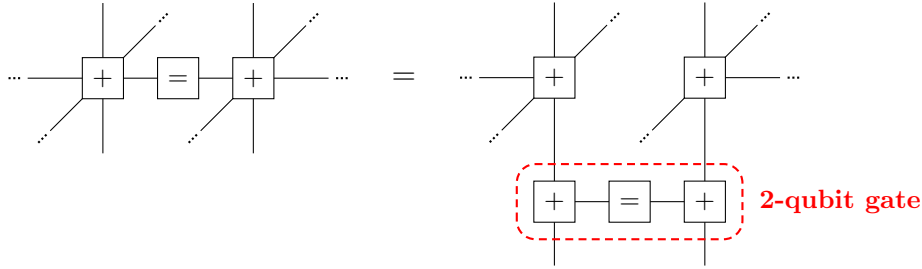

    \centering
    \include{figures/gate_splitting_special_case}
    \caption{Splitting off of two-qubit gates from a layer of the 3D tensor network for the point sector 3D surface code contraction.}
    \label{fig:splitting_special_case}
\end{figure}

The tensors on the bonds between two cubic lattice sites require a small modification in the contraction algorithm.
The tensors on the horizontal bonds can be integrated into the two-qubit gates of the simple update contraction procedure, whereas the tensors on vertical bonds can be exactly contracted in between layers.
Since the tensors on the sites are either $+$ or $=$ nodes, the splitting up of the layer into two-qubit gates can be done very naturally as depicted in~\Cref{fig:splitting_special_case}.

For our numerical experiments, we use a maximum bond dimension of 24 for the simple update contraction procedure and a maximum bond dimension of 32 for the MPS contraction of the final 2D tensor network.
The results are depicted in~\Cref{fig:threshold_z}.
Also depicted are results for the minimum-weight perfect matching decoder, which finds the most likely error and is thus the optimal decoder up to degeneracy.
The optimal threshold for the point sector is known to be around 3.3\%~\cite{ohno_phase_2004}.
We used PyMatching2~\cite{higgott_sparse_2023} for the matching implementation.

\subsection{Loop Sector of 3D Surface Code}
By the considerations in~\Cref{sec:css_special_case}, the detector network for the point sector (see~\Cref{fig:tn_pointsec}) and the generator network for the loop sector of the 3D surface code are essentially the same, except the $+$ and $=$ nodes are interchanged.
So due to the same reasoning as in the point sector, we choose to use the generator tensor network for the loop sector as it fits more naturally in a cubic form.
Since the tensor network is topologically the same as the one of the point sector, we use the analogous contraction technique: The gate splitting is identical as in~\Cref{fig:splitting_special_case}, except that the $=$ and $+$ nodes are interchanged.

For our numerical experiments, we use a maximum bond dimension of 24 for the simple update contraction procedure and a maximum bond dimension of 48 for the MPS contraction of the final 2D tensor network.
To our knowledge, the state-of-the-art decoding algorithm on the loop sector of the 3D surface code is BP+OSD as reported by Huang \emph{et al.}~\cite{quintavalle_single-shot_2021}, though they consider periodic boundary conditions. 
They report a threshold of around 21.55\%, while the optimal threshold is known~\cite{ozeki_multicritical_1998,hasenbusch_magnetic-glassy_2007} to be roughly 23.180\%.
Our numerical results for the TN decoder and BP+OSD (both on non-periodic boundary conditions) are depicted in~\Cref{fig:threshold_x}.
We employ Roffe \emph{et al.}'s BP+OSD implementation~\cite{roffe_decoding_2020,roffe_ldpc_2022} using the min-sum algorithm and combination-sweep OSD with an order of 60.

\subsection{Depolarizing Noise on 3D Surface Code}
For depolarizing noise we make use of the detector tensor network.
This means that now both the point-like and loop-like stabilizers are represented by nodes in the network.
This incurs an additional difficulty: In the standard description of the 3D surface code, the loop-like stabilizer generators are redundant.
Therefore, one must remove a subset of the loop stabilizers to obtain a linearly independent generator set.
For this purpose we remove all the stabilizers on the `top' of a cell, other than those on one (smooth) boundary.
The tensor network for $d=3$ is depicted in~\Cref{fig:tn_depol}.
Notice that compared to the point or loop sector, the lattice size of the resulting cubic tensor network is doubled, or respectively the volume is increased eight-fold.
Furthermore, not all vertices and edges of the cubic lattice are occupied.
In order to bring the tensor network to a truly cubic form amenable to our contracition algorithm, we fill up these empty slots with trivial tensors (having value $1$) and trivial bonds of dimension $1$. 

We use a maximum bond dimension of 20 for the simple update contraction procedure and a maximum bond dimension of 64 for the MPS contraction of the final 2D tensor network.
The maximum bond dimension of the singular value decomposition of the splitting technique (see~\Cref{fig:splitting}) is chosen to be $4$.

In~\Cref{fig:threshold_depol} we depict the numerical results comparing our TN decoder with the current state-of-the-art decoder, which to our knowledge is BP+OSD which has been reported to have a threshold of $5.95 \pm 0.03 \%$\cite{huang_tailoring_2022}.
Note that our BP+OSD results seem to perform significantly better than the threshold indicated by these authors---this could however just be boundary effects that disappear at larger distances.
We use Roffe \emph{et al.}'s BP+OSD implementation~\cite{roffe_decoding_2020,roffe_ldpc_2022} using the min-sum algorithm and combination-sweep OSD with an order of 60. As in \cite{huang_tailoring_2022}, phase-flips are decoded assuming i.i.d.\ phase-flip noise first (with the noise rate given by the marginal of depolarizing noise), and subsequently bit-flips are decoded, using the conditional distribution of bit-flips given the phase-flip pattern from the first step. 

We also attempted to realize a TN decoder using the generator tensor network, but it performed significantly worse than the detector TN decoder. 

\subsection{Circuit-level Noise}
We study the rotated surface code under circuit-level depolarizing noise.
More concretely, for a distance $d$ code we study the protocol consisting of $d$ repeated stabilizer measurement rounds.
The first $X$-stabilizer measurement round serves as an initialization of the logical qubit in the $X$ basis, and at the very end of the protocol all data qubits are measured in the $X$ basis to realize a logical $X$ measurement.
The goal of the decoder is to error-correct the outcome of said logical $X$ measurement.
After each gate, after each reset and before every measurement, we assume that the involved qubits undergo depolarizing noise (either 1-qubit or 2-qubit depolarizing noise accordingly) of some strength $p$.
We use the Stim package~\cite{gidney_stim_2021} to generate the DEM model, so we refer the readers to there for more information about the precise details about the considered circuit.

\begin{sidewaysfigure}[h]
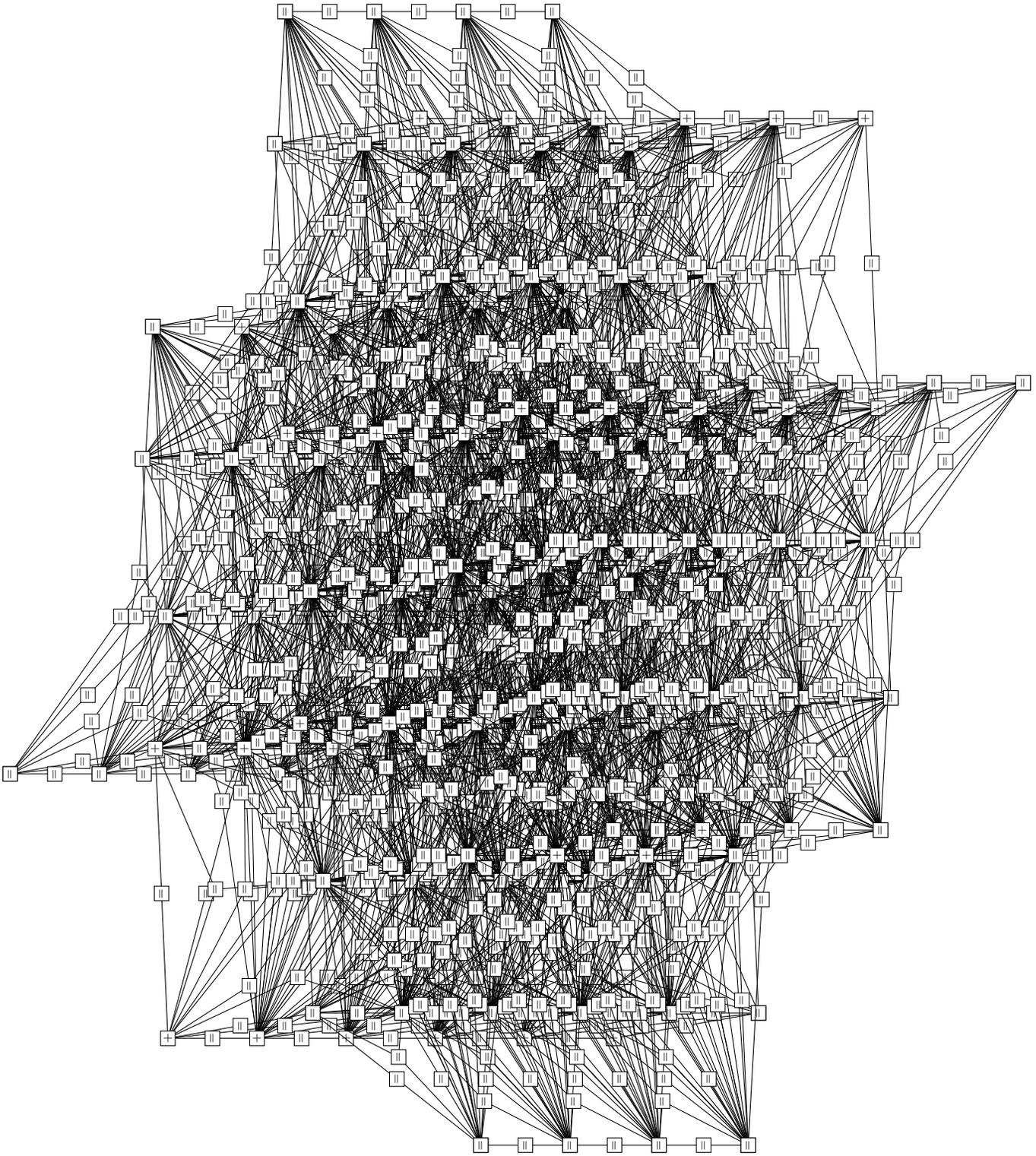

    \centering
    \tiny
    \include{figures/tn_circ_large}
    \normalsize
    \caption{Circuit-level noise tensor network derived from detector error model for the $d=5$ rotated surface code with $5$ rounds of measurements.}
    \label{fig:tn_circ_large}
\end{sidewaysfigure}

We use the detector picture to represent the decoding problem.
Therefore, following the discussion in~\Cref{sec:dem}, the $i$th error mechanism of the DEM is represented by a weighted equality node in the tensor network, i.e. a tensor that takes value $(1-p_i)$ if all incoming legs are $0$, value $p_i$ if all incoming legs are $1$ and value $0$ elsewise.
The $i$th detector is represented by a check tensor that takes the value $(1-m_i)$ if the parity of the incoming legs is $0$ and value $m_i$ elsewise.
An equality and check node are connected if and only if the involved error mechanism causes said detector to flip.
We use the Walsh-Hadamard transform discussed in~\Cref{sec:hadamard_trick} to realize the logical parity check.
For example,~\Cref{fig:tn_circ} depicts the circuit-level tensor network for the $d=3$ rotated surface code.
A depiction of the $d=5$ tensor network can be found at the end of the manuscript in~\Cref{fig:tn_circ_large}.
Clearly, the resulting tensor network is not even remotely close to a cubic structure: Some of the check nodes have a very high degree, which is caused by the high number of distinct error mechanisms that can cause said detector to flip.
Furthermore, the size of the tensor network is much higher than the non-circuit-level noise counterparts: The $d=3$, $d=5$ and $d=7$ tensor networks contain 245, 1799 and 6351 tensors respectively.

\begin{figure}[h]
    \centering
    \tiny
    \include{figures/tn_circ}
    \normalsize
    \caption{Circuit-level noise tensor network derived from detector error model for the $d=3$ rotated surface code with $3$ rounds of measurements. They weighting of the check and equality nodes is not depicted. The Hadamard nodes from the Walsh-Hadamard transform are also not depicted.}
    \label{fig:tn_circ}
\end{figure}

Clearly, the issues laid out above make it impossible to directly contract the circuit-level tensor network directly.
In fact, just na\"ively storing the tensors would require an infeasible amount of memory due to the high tensor degrees.
In order to address these problems, we propose the following scheme, in which the complicated DEM-derived tensor network is compressed and brought into a cubic form more amenable to our contraction scheme.

\begin{figure}[h]
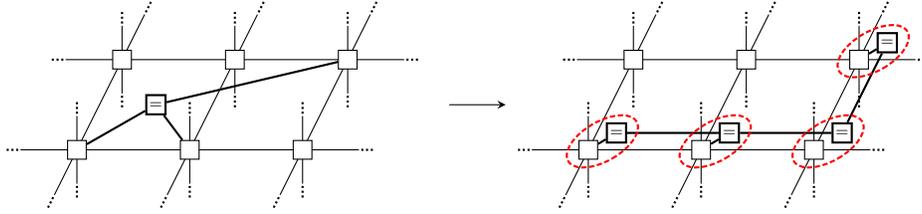

    \centering
    \include{figures/snaking}
    \caption{``Snaking'' of a weighted equality node corresponding to an error mechanism along the cubic TN. The red dashed lines depict the updated tensors of the cubic TN. Clearly, the dimensions of the bonds involved in the snaking are doubled.}
    \label{fig:snaking}
\end{figure}

First, only consider the check nodes (corresponding to the detectors) and arrange them in a cubic lattice according to their location in space-time.
Then, add the equality nodes (corresponding to the error mechanisms) one-by-one into this tensor network.
If this step was done without approximation, then the resulting tensor network would clearly be equivalent to the desired circuit-level tensor network.
To bring the equality node into something that respects the cubic lattice structure, we ``snake'' it along the lattice as depicted in~\Cref{fig:snaking}.
Clearly, this will cause the bond dimensions of the cubic tensor network to grow very quickly, so we use the simple update method to truncate the bonds to some maximum bond dimension.

During this compression step we keep an open edge on each site to represent the detector measurement outcome.
When a sample is to be decoded, we contract the corresponding observed detector values on each site of the cubic tensor network.
The resulting 3D tensor network can then be contracted with our technique described in~\Cref{sec:approxcontract}.
Thanks to this trick, the compression step needs only to be performed \emph{once} offline and not every time a new sample is to be decoded.
This largely alleviates the cost of this compression step and allows us to perform it with much larger bond dimensions.

\Cref{fig:threshold_circ} depicts the numerical results of our experiments comparing the tensor network decoder to the commonly used minimum-weight perfect matching decoder, implemented using PyMatching~\cite{higgott_pymatching_2022,higgott_sparse_2023}.
For the offline pre-compression of the circuit-level tensor network, we use a maximum bond dimension of 16.
Technically, if the physical error rate $p$ is varied, one would have to generate a separate pre-compressed tensor network for every value of $p$.
For convenience, we forgo this step and only generate one pre-compressed tensor network for a value of $p=1\%$.
Before being used for decoding (but after all error mechanisms have been compressed), we further truncate all bonds to a dimension of 8 in order to improve the speed of the decoder.
To contract the resulting 3D cubic tensor network, we use a maximum bond dimension of 12 for the simple update, a maximum bond dimension of 64 for the MPS contraction and maximum bond dimension of 12 for the splitting process.
For $d=7$, we increase the simple update bond dimension to 20 and the splitting bond dimension to 14.

\section{Discussion}
\label{sec:discussion}
Decoding 3D codes is a notoriously difficult problem, especially when the code does not allow for a matching-like decoding algorithm.
We propose a new decoding scheme that approximates the optimal decoder and outperforms the current state-of-the-art algorithm on the point sector, loop sector, and depolarizing noise for 3D surface codes of moderate size.
Once the size of the code becomes too large, typically around linear size (code distance) $d=11$, our 3D tensor network decoder starts facing numerical issues, which degrades the accuracy. 
This is in stark contrast to the 2D tensor network decoder (based on MPS truncation) which ``out of the box'' is both fast and reliable even for large codes.

This can be most likely attributed to the lack of a canonical form which cannot be defined for a PEPS in contrast to an MPS, making it difficult to realize bond truncations accurately.
There are many techniques and heuristics for dealing with this issue.
While we roughly explored several of them during the development of this paper, we settled for the simple update technique, since in our tests it provided the best accuracy-to-speed ratio.
We experimented with bringing the PEPS into the Vidal gauge before truncation, either with belief propagation or with repeated trivial simple update steps, but the convergence of the gauging procedure was very slow and made up for the vast majority of the contraction time.
We also investigated exactly contracting a complete layer at once before truncating the resulting bonds, but again the accuracy-time trade-off was not favorable.

The space of possibilities to realize approximate 3D tensor network contraction is much larger than in 2D, and is still an active field of research. Certainly the possibilities greatly exceed our limited exploration. 
We therefore expect that future work will be able to improve upon our results.

The 3D surface code can in some sense be considered one of the simplest 3D codes to decode, since the tensor network structure is naturally (almost) cubic.
To show that our 3D contraction algorithm does not necessarily rely on a code that induces such a nicely structured tensor network, we also demonstrate our technique for decoding circuit-level noise, which arguably exhibits the most complicated structure encountered in practice.
We show that our technique outperforms the matching decoder, though we did not compare it to the state-of-the-art algorithm belief-matching, which currently does not have a publicly available implementation.
Our technique is not very optimized, but rather serves the purpose of demonstrating that the general TN decoding method works in principle.
We fully expect that more careful and tedious tuning of the decoder parameters could significantly improve the performance and/or accuracy of the TN decoder on circuit-level noise.
Similarly, we spent very little time with trying out different snaking and truncation procedures for the tensor network pre-compression step.
Further research in optimizing the decoder and comparing it with other circuit-level decoders is necessary.
The possible impact for near-term quantum error correction experiments is very high, since the involved distances are rather small and the decoding can typically be performed offline.
So a very accurate-but-slow tensor network decoder could facilitate the demonstration of break-even error correction.

Another point that requires further investigation is whether the detector or generator picture performs better in practice.
While we did not test this systematically, we did notice that the detector tensor network seems much easier to contract for depolarizing noise compared to the generator tensor network.
Also for the circuit-level noise one could make use of the generator picture tensor network, though this would require first finding a dual parity check matrix to the detector error model.

\section*{Acknowledgements}
This work was supported by the ETH Quantum Center, the National
Centres for Competence in Research in Quantum Science
and Technology (QSIT) and The Mathematics of Physics
(SwissMAP), and the Swiss National Science Foundation Sinergia grant CRSII5\_186364.
The authors are grateful to Michael Vasmer for useful discussions.
We extensively used the panqec library\cite{huang_panqec_2023} and thank Eric Huang and Arthur Pesah for their insights.

\printbibliography[heading=bibintoc,title={\large References}]


\newpage

\end{document}

%% file: figures/simpleTN.tex
\pgfmathsetmacro\MathAxis{height("$\vcenter{}$")}
\begin{tikzpicture}[
baseline={(0, -0.25cm-\MathAxis pt)},
square/.style={regular polygon,regular polygon sides=4,inner sep=0,minimum height=15},
scale=0.8
]

\node[square,draw] (f1) at (-2,0) {$f_1$};
\node[square,draw] (f2) at (2,0) {$f_2$};
\node[square,draw] (f3) at (0,-1) {$f_3$};
\node[square] (x1) at (-4,-0) {$x_1$};
\node[square] (x2) at (4,0) {$x_2$};
\draw (x1) -- (f1) -- node[midway,below left] {$y_1$} (f3) -- node[midway,below right] {$y_2$} (f2) -- (x2);
\draw (f1) -- node[midway,above] {$y_3$} (f2);

\end{tikzpicture}

%% file: figures/simpleTN2.tex
\pgfmathsetmacro\MathAxis{height("$\vcenter{}$")}
\begin{tikzpicture}[
baseline={(0, -0.25cm-\MathAxis pt)},
square/.style={regular polygon,regular polygon sides=4,inner sep=0,minimum height=15},
scale=0.8
]

\node[square,draw] (f1) at (-2,0) {$f_1$};
\node[square,draw] (f2) at (2,0) {$f_2$};
\node[square,draw] (f3) at (0,-1) {$f_3$};
\node[square] (x1) at (-4,-0) {\phantom{$x_1$}};
\node[square] (x2) at (4,0) {\phantom{$x_2$}};
\draw (x1) -- node[midway,below] {$x_1$} (f1) -- node[midway,below left] {$y_1$} (f3) -- node[midway,below right] {$y_2$} (f2) -- node[midway,below] {$x_2$} (x2);
\draw (f1) -- node[midway,above] {$y_3$} (f2);

\end{tikzpicture}

%% file: figures/simpleTN3.tex
\pgfmathsetmacro\MathAxis{height("$\vcenter{}$")}
\begin{tikzpicture}[
baseline={(0, -0.375cm-\MathAxis pt)},
square/.style={regular polygon,regular polygon sides=4,inner sep=0,minimum height=15},
scale=0.8
]

\node[square,draw] (f1) at (-2,0) {$f_1$};
\node[square,draw] (f2) at (2,0) {$f_2$};
\node[square,draw] (f3) at (0,-1) {$f_3$};
\node[square,draw] (y3) at (0,0) {\eq};
\node[square] (x1) at (-4,-0) {\phantom{$x_1$}};
\node[square] (x2) at (4,0) {\phantom{$x_2$}};
\draw (x1) -- node[midway,below] {$x_1$} (f1) -- node[midway,below left] {$y_1$} (f3) -- node[midway,below right] {$y_2$} (f2) -- node[midway,below] {$x_2$} (x2);
\draw (f1) --  (y3) -- (f2);
\draw (y3) -- (f3);

\end{tikzpicture}

%% file: figures/tn_detector_col.tex
\begin{tikzpicture}[square/.style={regular polygon,regular polygon sides=4,inner sep=0,minimum height=15},scale=0.8]

\foreach \i in {1,2,3,4,5} {
    \node[square,draw,fill=red!20] (z\i) at (3.5*\i,0) {\eq};
    \node[square,draw,fill=blue!20] (x\i) at (3.5*\i+1,0) {\eq};
    \node[minimum height=15,minimum width=32,draw,fill=white] (prob\i) at (3.5*\i+0.5,-1) {\small $P_{{\i}}$};
}

\foreach \i in {1,2,3,4,5,6} {
    \node[square,draw,fill=white] (p\i) at (3*\i+0.5,4.2+0.75) {$+$};
}
\foreach \i in {1,2,3,4} {
    \node (sy\i) at (3*\i+0.5,5.2+0.75) {$m_\i$};
    \draw (sy\i) -- (p\i);
}

\node (sy5) at (15.5,5.2+0.75) {$b$};
\node (sy6) at (18.5,5.2+0.75) {$a$};
\draw (sy5) -- (p5);
\draw (sy6) -- (p6);

\draw[blue]
	(p5) -- (x1)	
	(p6) -- (x4)
	(p6) -- (x5);
\draw[red] 
	(p5) -- (z2)
	(p6) -- (z2)
	(p5) -- (z4);

\draw[blue] 
	(p1) -- (x1) 
	(p1) -- (x4)
	(p2) -- (x2)
	(p2) -- (x5)
	(p3) -- (x1)
	(p3) -- (x3)
	(p4) -- (x2)
	(p4) -- (x4);

\draw[red] 
	(p1) -- (z2)
	(p1) -- (z3)
	(p2) -- (z3)
	(p2) -- (z4)
	(p3) -- (z4)
	(p3) -- (z5)
	(p4) -- (z1)
	(p4) -- (z5);

\foreach \i in {1,2,3,4,5} {
	\draw[blue] (x\i) -- (prob\i.north-|x\i);
	\draw[red] (z\i) -- (prob\i.north-|z\i);
}

\end{tikzpicture}

%% file: figures/Hadamard.tex
\pgfmathsetmacro\MathAxis{height("$\vcenter{}$")}
\begin{tikzpicture}[
baseline={(0, -\MathAxis pt)},
square/.style={regular polygon,regular polygon sides=4,inner sep=0,minimum height=15},
scale=0.8
]

\node at (0,0) {$\propto$};
\begin{scope}[xshift=-25mm]
\draw[rounded corners=1] (-1.75,0) -- (0,0) |- (1.75,0.5);
\draw[rounded corners=1] (0,0) |- (1.75,-0.5);
\node[square,draw,fill=white] at (0,0) {\eq};
\node[square,draw,fill=white] at (1,0.5) {\small $H$};
\node[square,draw,fill=white] at (1,-0.5) {\small $H$};
\node[square,draw,fill=white] at (-1,0) {\small $H$};
\end{scope}
\begin{scope}[xshift=15mm]
\draw[rounded corners=1] (-0.75,0) -- (0,0) |- (0.75,0.5);
\draw[rounded corners=1] (0,0) |- (0.75,-0.5);

\node[square,draw,fill=white] at (0,0) {$+$};
\end{scope}

\end{tikzpicture}

%% file: figures/WalshHadamard.tex
\pgfmathsetmacro\MathAxis{height("$\vcenter{}$")}
\begin{tikzpicture}[
baseline={(0, -\MathAxis pt)},
square/.style={regular polygon,regular polygon sides=4,inner sep=0,minimum height=15},
scale=0.8
]
\def\had{
\draw[rounded corners=1] (-1.75,0) -- (0,0) |- (1.75,0.5);
\draw[rounded corners=1] (0,0) |- (1.75,-0.5);
\node[square,draw,fill=white] at (0,0) {\eq};
\node[square,draw,fill=white] at (1,0.5) {\small $H$};
\node[square,draw,fill=white] at (1,-0.5) {\small $H$};
\node[square,draw,fill=white] at (-1,0) {\small $H$};
}
\begin{scope}[xshift=-26mm]
\had
\end{scope}
\begin{scope}[xshift=26mm,xscale=-1]
\had
\end{scope}
\node[cloud, fill=white,draw, minimum height=19mm,minimum width=17mm,cloud puffs=11,cloud ignores aspect,cloud puff arc=120] (c) at (0,0) {};
\draw[rounded corners,dashed] (-3.0,-1.5) rectangle (3.0,1.5);

\end{tikzpicture}

%% file: figures/tn_gauge_col.tex
\begin{tikzpicture}[square/.style={regular polygon,regular polygon sides=4,inner sep=0,minimum height=15},scale=0.8]

\foreach \i in {1,2,3,4,5} {
  \node[square,draw,fill=red!20] (x\i) at (3.5*\i,0) {$+$};
  \node[red,inner sep=1pt] (rx\i) at (3.5*\i-0.75,0) {\small $r^x_{\i}$};
  \draw[red] (rx\i) -- (x\i);
  \node[square,draw,fill=blue!20] (z\i) at (3.5*\i+1,0) {$+$};
  \node[blue,inner sep=1pt] (rz\i) at (3.5*\i+1.75,0) {\small $r^z_{\i}$};
  \draw[blue] (rz\i) -- (z\i);
  \node[minimum height=15,minimum width=32,draw,fill=white] (p\i) at (3.5*\i+0.5,-1) {\small $P_{{\i}}$};
  }

\foreach \i in {1,2,3,4} {
  \node[square,draw,fill=white] (s\i) at (3.5*\i+2.25,4) {\eq};
}

\draw[red]
	(s1) -- (x1)
	(s1) -- (x4)
	(s2) -- (x2)
	(s2) -- (x5)
	(s3) -- (x1)
	(s3) -- (x3)
	(s4) -- (x2)
	(s4) -- (x4);

\draw[blue]
	(s1) -- (z2)
	(s1) -- (z3)
	(s2) -- (z3)
	(s2) -- (z4)
	(s3) -- (z4)
	(s3) -- (z5)
	(s4) -- (z1)
	(s4) -- (z5);

\foreach \i in {1,2,3,4,5} {
	\draw[red] (x\i) -- (p\i.north-|x\i);
	\draw[blue] (z\i) -- (p\i.north-|z\i);
}

\end{tikzpicture}

%% file: figures/sc2d_detector_col.tex
\begin{tikzpicture}[very thick,scale=1.8]
\tikzstyle{tnnode}=[regular polygon,regular polygon sides=4,draw=black,fill=white,inner sep=0,outer sep=0]
\draw[red] 
	(2,0) -- (1.0,-0.3)
	(2,0) -- (3.0,-0.3)
	(2,0) -- (2.0,0.7)
	(2,2) -- (1.0,1.7)
	(2,2) -- (3.0,1.7)
	(2,2) -- (2.0,0.7)
	(2,2) -- (2.0,2.7)
	(2,4) -- (1.0,3.7)
	(2,4) -- (3.0,3.7)
	(2,4) -- (2.0,2.7)
	(4,0) -- (3.0,-0.3)
	(4,0) -- (5.0,-0.3)
	(4,0) -- (4.0,0.7)
	(4,2) -- (3.0,1.7)
	(4,2) -- (5.0,1.7)
	(4,2) -- (4.0,0.7)
	(4,2) -- (4.0,2.7)
	(4,4) -- (3.0,3.7)
	(4,4) -- (5.0,3.7)
	(4,4) -- (4.0,2.7);

\draw[blue]
	(1,1) -- (1.79,1.0)
	(1,1) -- (0.79,0.0)
	(1,1) -- (0.79,2.0)
	(1,3) -- (1.79,3.0)
	(1,3) -- (0.79,2.0)
	(1,3) -- (0.79,4.0)
	(3,1) -- (1.79,1.0)
	(3,1) -- (3.79,1.0)
	(3,1) -- (2.79,0.0)
	(3,1) -- (2.79,2.0)
	(3,3) -- (1.79,3.0)
	(3,3) -- (3.79,3.0)
	(3,3) -- (2.79,2.0)
	(3,3) -- (2.79,4.0)
	(5,1) -- (3.79,1.0)
	(5,1) -- (4.79,0.0)
	(5,1) -- (4.79,2.0)
	(5,3) -- (3.79,3.0)
	(5,3) -- (4.79,2.0)
	(5,3) -- (4.79,4.0);

\draw [blue] 
	(0.79,0.0) -- (1.25,0.25)
	(0.79,2.0) -- (1.25,2.25)
	(0.79,4.0) -- (1.25,4.25)
	(2.79,0.0) -- (3.25,0.25)
	(2.79,2.0) -- (3.25,2.25)
	(2.79,4.0) -- (3.25,4.25)
	(4.79,0.0) -- (5.25,0.25)
	(4.79,2.0) -- (5.25,2.25)
	(4.79,4.0) -- (5.25,4.25)
	(1.79,1.0) -- (2.25,1.25)
	(1.79,3.0) -- (2.25,3.25)
	(3.79,1.0) -- (4.25,1.25)
	(3.79,3.0) -- (4.25,3.25);

\draw[red]
	(1.0,-0.3) -- (1.25,0.25)
	(1.0,1.7) -- (1.25,2.25)
	(1.0,3.7) -- (1.25,4.25)
	(3.0,-0.3) -- (3.25,0.25)
	(3.0,1.7) -- (3.25,2.25)
	(3.0,3.7) -- (3.25,4.25)
	(5.0,-0.3) -- (5.25,0.25)
	(5.0,1.7) -- (5.25,2.25)
	(5.0,3.7) -- (5.25,4.25)
	(2.0,0.7) -- (2.25,1.25)
	(2.0,2.7) -- (2.25,3.25)
	(4.0,0.7) -- (4.25,1.25)
	(4.0,2.7) -- (4.25,3.25);

\node (1_0_X) at (0.79,0.0) [tnnode,fill=blue!20,inner sep=-1pt] {\eq};
\node (1_0_Z) at (1.0,-0.3) [tnnode,fill=red!20,inner sep=-1pt] {\eq};
\node (1_0_P) at (1.25,0.25) [tnnode] {$P_{1}$};
\draw[red] (1_0_Z.west) -- (0.0,-0.3);
\node (1_0_H) at (0.4,-0.3) [tnnode] {$H$};
\node (1_0_L) at (0.0,-0.3) [red,tnnode,draw=none] {$b$};
\draw[blue] (1_0_X.south) -- (0.79,-1.0);
\node (1_0_H) at (0.79,0.0-0.6) [tnnode] {$H$};
\node (1_0_L) at (0.79,0.0-1.0) [blue,tnnode,draw=none] {$a$};
\node (1_2_X) at (0.79,2.0) [tnnode,fill=blue!20,inner sep=-1pt] {\eq};
\node (1_2_Z) at (1.0,1.7) [tnnode,fill=red!20,inner sep=-1pt] {\eq};
\node (1_2_P) at (1.25,2.25) [tnnode] {$P_{2}$};
\draw[red] (1_2_Z.west) -- (0.0,1.7);
\node (1_2_H) at (0.4,1.7) [tnnode] {$H$};
\node (1_2_L) at (0.0,1.7) [red,tnnode,draw=none] {$b$};
\node (1_4_X) at (0.79,4.0) [tnnode,fill=blue!20,inner sep=-1pt] {\eq};
\node (1_4_Z) at (1.0,3.7) [tnnode,fill=red!20,inner sep=-1pt] {\eq};
\node (1_4_P) at (1.25,4.25) [tnnode] {$P_{3}$};
\draw[red] (1_4_Z.west) -- (0.0,3.7);
\node (1_4_H) at (0.4,3.7) [tnnode] {$H$};
\node (1_4_L) at (0.0,3.7) [red,tnnode,draw=none] {$b$};
\node (3_0_X) at (2.79,0.0) [tnnode,fill=blue!20,inner sep=-1pt] {\eq};
\node (3_0_Z) at (3.0,-0.3) [tnnode,fill=red!20,inner sep=-1pt] {\eq};
\node (3_0_P) at (3.25,0.25) [tnnode] {$P_{6}$};
\draw[blue] (3_0_X.south) -- (2.79,-1.0);
\node (3_0_H) at (2.79,0.0-0.6) [tnnode] {$H$};
\node (3_0_L) at (2.79,0.0-1.0) [blue,tnnode,draw=none] {$a$};
\node (3_2_X) at (2.79,2.0) [tnnode,fill=blue!20,inner sep=-1pt] {\eq};
\node (3_2_Z) at (3.0,1.7) [tnnode,fill=red!20,inner sep=-1pt] {\eq};
\node (3_2_P) at (3.25,2.25) [tnnode] {$P_{7}$};
\node (3_4_X) at (2.79,4.0) [tnnode,fill=blue!20,inner sep=-1pt] {\eq};
\node (3_4_Z) at (3.0,3.7) [tnnode,fill=red!20,inner sep=-1pt] {\eq};
\node (3_4_P) at (3.25,4.25) [tnnode] {$P_{8}$};
\node (5_0_X) at (4.79,0.0) [tnnode,fill=blue!20,inner sep=-1pt] {\eq};
\node (5_0_Z) at (5.0,-0.3) [tnnode,fill=red!20,inner sep=-1pt] {\eq};
\node (5_0_P) at (5.25,0.25) [tnnode] {$P_{11}$};
\draw[blue] (5_0_X.south) -- (4.79,-1.0);
\node (5_0_H) at (4.79,0.0-0.6) [tnnode] {$H$};
\node (5_0_L) at (4.79,0.0-1.0) [blue,tnnode,draw=none] {$a$};
\node (5_2_X) at (4.79,2.0) [tnnode,fill=blue!20,inner sep=-1pt] {\eq};
\node (5_2_Z) at (5.0,1.7) [tnnode,fill=red!20,inner sep=-1pt] {\eq};
\node (5_2_P) at (5.25,2.25) [tnnode] {$P_{12}$};
\node (5_4_X) at (4.79,4.0) [tnnode,fill=blue!20,inner sep=-1pt] {\eq};
\node (5_4_Z) at (5.0,3.7) [tnnode,fill=red!20,inner sep=-1pt] {\eq};
\node (5_4_P) at (5.25,4.25) [tnnode] {$P_{13}$};
\node (2_1_X) at (1.79,1.0) [tnnode,fill=blue!20,inner sep=-1pt] {\eq};
\node (2_1_Z) at (2.0,0.7) [tnnode,fill=red!20,inner sep=-1pt] {\eq};
\node (2_1_P) at (2.25,1.25) [tnnode] {$P_{4}$};
\node (2_3_X) at (1.79,3.0) [tnnode,fill=blue!20,inner sep=-1pt] {\eq};
\node (2_3_Z) at (2.0,2.7) [tnnode,fill=red!20,inner sep=-1pt] {\eq};
\node (2_3_P) at (2.25,3.25) [tnnode] {$P_{5}$};
\node (4_1_X) at (3.79,1.0) [tnnode,fill=blue!20,inner sep=-1pt] {\eq};
\node (4_1_Z) at (4.0,0.7) [tnnode,fill=red!20,inner sep=-1pt] {\eq};
\node (4_1_P) at (4.25,1.25) [tnnode] {$P_{9}$};
\node (4_3_X) at (3.79,3.0) [tnnode,fill=blue!20,inner sep=-1pt] {\eq};
\node (4_3_Z) at (4.0,2.7) [tnnode,fill=red!20,inner sep=-1pt] {\eq};
\node (4_3_P) at (4.25,3.25) [tnnode] {$P_{10}$};
\node (2_0) at (2,0) [tnnode] {$+_{m_{3}}$};
\node (2_2) at (2,2) [tnnode] {$+_{m_{4}}$};
\node (2_4) at (2,4) [tnnode] {$+_{m_{5}}$};
\node (4_0) at (4,0) [tnnode] {$+_{m_{8}}$};
\node (4_2) at (4,2) [tnnode] {$+_{m_{9}}$};
\node (4_4) at (4,4) [tnnode] {$+_{m_{10}}$};
\node (1_1) at (1,1) [tnnode] {$+_{m_{1}}$};
\node (1_3) at (1,3) [tnnode] {$+_{m_{2}}$};
\node (3_1) at (3,1) [tnnode] {$+_{m_{6}}$};
\node (3_3) at (3,3) [tnnode] {$+_{m_{7}}$};
\node (5_1) at (5,1) [tnnode] {$+_{m_{11}}$};
\node (5_3) at (5,3) [tnnode] {$+_{m_{12}}$};
\end{tikzpicture}

%% file: figures/sc2d_generator_col.tex
\begin{tikzpicture}[very thick,scale=2]
\tikzstyle{tnnode}=[regular polygon,regular polygon sides=4,draw=black,fill=white,inner sep=0,outer sep=0]

\draw[blue]
	(2,0) -- (1.0,-0.4)
	(2,0) -- (3.0,-0.4)
	(2,0) -- (2.0,0.6)
	(2,2) -- (1.0,1.6)
	(2,2) -- (3.0,1.6)
	(2,2) -- (2.0,0.6)
	(2,2) -- (2.0,2.6)
	(2,4) -- (1.0,3.6)
	(2,4) -- (3.0,3.6)
	(2,4) -- (2.0,2.6)
	(4,0) -- (3.0,-0.4)
	(4,0) -- (5.0,-0.4)
	(4,0) -- (4.0,0.6)
	(4,2) -- (3.0,1.6)
	(4,2) -- (5.0,1.6)
	(4,2) -- (4.0,0.6)
	(4,2) -- (4.0,2.6)
	(4,4) -- (3.0,3.6)
	(4,4) -- (5.0,3.6)
	(4,4) -- (4.0,2.6);

\draw[red]
	(1,1) -- (1.69,1.0)
	(1,1) -- (0.69,0.0)
	(1,1) -- (0.69,2.0)
	(1,3) -- (1.69,3.0)
	(1,3) -- (0.69,2.0)
	(1,3) -- (0.69,4.0)
	(3,1) -- (1.69,1.0)
	(3,1) -- (3.69,1.0)
	(3,1) -- (2.69,0.0)
	(3,1) -- (2.69,2.0)
	(3,3) -- (1.69,3.0)
	(3,3) -- (3.69,3.0)
	(3,3) -- (2.69,2.0)
	(3,3) -- (2.69,4.0)
	(5,1) -- (3.69,1.0)
	(5,1) -- (4.69,0.0)
	(5,1) -- (4.69,2.0)
	(5,3) -- (3.69,3.0)
	(5,3) -- (4.69,2.0)
	(5,3) -- (4.69,4.0);

\draw[red]
	(0.69,0.0) -- (1.25,0.25)
	(0.69,2.0) -- (1.25,2.25)
	(0.69,4.0) -- (1.25,4.25)
	(2.69,0.0) -- (3.25,0.25)
	(2.69,2.0) -- (3.25,2.25)
	(2.69,4.0) -- (3.25,4.25)
	(4.69,0.0) -- (5.25,0.25)
	(4.69,2.0) -- (5.25,2.25)
	(4.69,4.0) -- (5.25,4.25)
	(1.69,1.0) -- (2.25,1.25)
	(1.69,3.0) -- (2.25,3.25)
	(3.69,1.0) -- (4.25,1.25)
	(3.69,3.0) -- (4.25,3.25);

\draw[blue]
	(1.0,-0.4) -- (1.25,0.25)
	(1.0,1.6) -- (1.25,2.25)
	(1.0,3.6) -- (1.25,4.25)
	(3.0,-0.4) -- (3.25,0.25)
	(3.0,1.6) -- (3.25,2.25)
	(3.0,3.6) -- (3.25,4.25)
	(5.0,-0.4) -- (5.25,0.25)
	(5.0,1.6) -- (5.25,2.25)
	(5.0,3.6) -- (5.25,4.25)
	(2.0,0.6) -- (2.25,1.25)
	(2.0,2.6) -- (2.25,3.25)
	(4.0,0.6) -- (4.25,1.25)
	(4.0,2.6) -- (4.25,3.25);

\node (1_0_X) at (0.69,0.0) [tnnode,fill=red!20,inner sep=-1pt] {$+_{r_{1}^{x}}$};
\node (1_0_Z) at (1.0,-0.4) [tnnode,fill=blue!20,inner sep=-1pt] {$+_{r_{1}^{z}}$};
\node (1_0_P) at (1.25,0.25) [tnnode] {$P_{1}$};
\node (1_2_X) at (0.69,2.0) [tnnode,fill=red!20,inner sep=-1pt] {$+_{r_{2}^{x}}$};
\node (1_2_Z) at (1.0,1.6) [tnnode,fill=blue!20,inner sep=-1pt] {$+_{r_{2}^{z}}$};
\node (1_2_P) at (1.25,2.25) [tnnode] {$P_{2}$};
\node (1_4_X) at (0.69,4.0) [tnnode,fill=red!20,inner sep=-1pt] {$+_{r_{3}^{x}}$};
\node (1_4_Z) at (1.0,3.6) [tnnode,fill=blue!20,inner sep=-1pt] {$+_{r_{3}^{z}}$};
\node (1_4_P) at (1.25,4.25) [tnnode] {$P_{3}$};
\node (3_0_X) at (2.69,0.0) [tnnode,fill=red!20,inner sep=-1pt] {$+_{r_{6}^{x}}$};
\node (3_0_Z) at (3.0,-0.4) [tnnode,fill=blue!20,inner sep=-1pt] {$+_{r_{6}^{z}}$};
\node (3_0_P) at (3.25,0.25) [tnnode] {$P_{6}$};
\node (3_2_X) at (2.69,2.0) [tnnode,fill=red!20,inner sep=-1pt] {$+_{r_{7}^{x}}$};
\node (3_2_Z) at (3.0,1.6) [tnnode,fill=blue!20,inner sep=-1pt] {$+_{r_{7}^{z}}$};
\node (3_2_P) at (3.25,2.25) [tnnode] {$P_{7}$};
\node (3_4_X) at (2.69,4.0) [tnnode,fill=red!20,inner sep=-1pt] {$+_{r_{8}^{x}}$};
\node (3_4_Z) at (3.0,3.6) [tnnode,fill=blue!20,inner sep=-1pt] {$+_{r_{8}^{z}}$};
\node (3_4_P) at (3.25,4.25) [tnnode] {$P_{8}$};
\node (5_0_X) at (4.69,0.0) [tnnode,fill=red!20,inner sep=-1pt] {$+_{r_{11}^{x}}$};
\node (5_0_Z) at (5.0,-0.4) [tnnode,fill=blue!20,inner sep=-1pt] {$+_{r_{11}^{z}}$};
\node (5_0_P) at (5.25,0.25) [tnnode] {$P_{11}$};
\node (5_2_X) at (4.69,2.0) [tnnode,fill=red!20,inner sep=-1pt] {$+_{r_{12}^{x}}$};
\node (5_2_Z) at (5.0,1.6) [tnnode,fill=blue!20,inner sep=-1pt] {$+_{r_{12}^{z}}$};
\node (5_2_P) at (5.25,2.25) [tnnode] {$P_{12}$};
\node (5_4_X) at (4.69,4.0) [tnnode,fill=red!20,inner sep=-1pt] {$+_{r_{13}^{x}}$};
\node (5_4_Z) at (5.0,3.6) [tnnode,fill=blue!20,inner sep=-1pt] {$+_{r_{13}^{z}}$};
\node (5_4_P) at (5.25,4.25) [tnnode] {$P_{13}$};
\node (2_1_X) at (1.69,1.0) [tnnode,fill=red!20,inner sep=-1pt] {$+_{r_{4}^{x}}$};
\node (2_1_Z) at (2.0,0.6) [tnnode,fill=blue!20,inner sep=-1pt] {$+_{r_{4}^{z}}$};
\node (2_1_P) at (2.25,1.25) [tnnode] {$P_{4}$};
\node (2_3_X) at (1.69,3.0) [tnnode,fill=red!20,inner sep=-1pt] {$+_{r_{5}^{x}}$};
\node (2_3_Z) at (2.0,2.6) [tnnode,fill=blue!20,inner sep=-1pt] {$+_{r_{5}^{z}}$};
\node (2_3_P) at (2.25,3.25) [tnnode] {$P_{5}$};
\node (4_1_X) at (3.69,1.0) [tnnode,fill=red!20,inner sep=-1pt] {$+_{r_{9}^{x}}$};
\node (4_1_Z) at (4.0,0.6) [tnnode,fill=blue!20,inner sep=-1pt] {$+_{r_{9}^{z}}$};
\node (4_1_P) at (4.25,1.25) [tnnode] {$P_{9}$};
\node (4_3_X) at (3.69,3.0) [tnnode,fill=red!20,inner sep=-1pt] {$+_{r_{10}^{x}}$};
\node (4_3_Z) at (4.0,2.6) [tnnode,fill=blue!20,inner sep=-1pt] {$+_{r_{10}^{z}}$};
\node (4_3_P) at (4.25,3.25) [tnnode] {$P_{10}$};
\node (2_0) at (2,0) [tnnode] {$=$};
\node (2_2) at (2,2) [tnnode] {$=$};
\node (2_4) at (2,4) [tnnode] {$=$};
\node (4_0) at (4,0) [tnnode] {$=$};
\node (4_2) at (4,2) [tnnode] {$=$};
\node (4_4) at (4,4) [tnnode] {$=$};
\node (1_1) at (1,1) [tnnode] {$=$};
\node (1_3) at (1,3) [tnnode] {$=$};
\node (3_1) at (3,1) [tnnode] {$=$};
\node (3_3) at (3,3) [tnnode] {$=$};
\node (5_1) at (5,1) [tnnode] {$=$};
\node (5_3) at (5,3) [tnnode] {$=$};
\end{tikzpicture}

%% file: figures/sc2d_bitflip_detector.tex
\begin{tikzpicture}[very thick,scale=1]
\tikzstyle{tnnode}=[regular polygon,regular polygon sides=4,draw=black,fill=white,inner sep=0,outer sep=0]
\draw (2,0) -- (1,0);
\draw (2,0) -- (3,0);
\draw (2,0) -- (2,1);
\draw (2,2) -- (1,2);
\draw (2,2) -- (3,2);
\draw (2,2) -- (2,1);
\draw (2,2) -- (2,3);
\draw (2,4) -- (1,4);
\draw (2,4) -- (3,4);
\draw (2,4) -- (2,3);
\draw (4,0) -- (3,0);
\draw (4,0) -- (5,0);
\draw (4,0) -- (4,1);
\draw (4,2) -- (3,2);
\draw (4,2) -- (5,2);
\draw (4,2) -- (4,1);
\draw (4,2) -- (4,3);
\draw (4,4) -- (3,4);
\draw (4,4) -- (5,4);
\draw (4,4) -- (4,3);
\node (1_0) at (1,0) [tnnode,inner sep=-1pt] {$=_p$};
\draw (1_0.west) -- (-0.30000000000000004,0);
\node (1_0_L) at (0.30000000000000004,0) [tnnode] {$H$};
\node (1_0_L) at (-0.30000000000000004,0) [tnnode,draw=none] {$b$};
\node (1_2) at (1,2) [tnnode,inner sep=-1pt] {$=_p$};
\draw (1_2.west) -- (-0.30000000000000004,2);
\node (1_2_L) at (0.30000000000000004,2) [tnnode] {$H$};
\node (1_2_L) at (-0.30000000000000004,2) [tnnode,draw=none] {$b$};
\node (1_4) at (1,4) [tnnode,inner sep=-1pt] {$=_p$};
\draw (1_4.west) -- (-0.30000000000000004,4);
\node (1_4_L) at (0.30000000000000004,4) [tnnode] {$H$};
\node (1_4_L) at (-0.30000000000000004,4) [tnnode,draw=none] {$b$};
\node (3_0) at (3,0) [tnnode,inner sep=-1pt] {$=_p$};
\node (3_2) at (3,2) [tnnode,inner sep=-1pt] {$=_p$};
\node (3_4) at (3,4) [tnnode,inner sep=-1pt] {$=_p$};
\node (5_0) at (5,0) [tnnode,inner sep=-1pt] {$=_p$};
\node (5_2) at (5,2) [tnnode,inner sep=-1pt] {$=_p$};
\node (5_4) at (5,4) [tnnode,inner sep=-1pt] {$=_p$};
\node (2_1) at (2,1) [tnnode,inner sep=-1pt] {$=_p$};
\node (2_3) at (2,3) [tnnode,inner sep=-1pt] {$=_p$};
\node (4_1) at (4,1) [tnnode,inner sep=-1pt] {$=_p$};
\node (4_3) at (4,3) [tnnode,inner sep=-1pt] {$=_p$};
\node (2_0) at (2,0) [tnnode, minimum width=0.7cm, max width=0.7cm] {$+_{m_{1}}$};
\node (2_2) at (2,2) [tnnode, minimum width=0.7cm, max width=0.7cm] {$+_{m_{2}}$};
\node (2_4) at (2,4) [tnnode, minimum width=0.7cm, max width=0.7cm] {$+_{m_{3}}$};
\node (4_0) at (4,0) [tnnode, minimum width=0.7cm, max width=0.7cm] {$+_{m_{4}}$};
\node (4_2) at (4,2) [tnnode, minimum width=0.7cm, max width=0.7cm] {$+_{m_{5}}$};
\node (4_4) at (4,4) [tnnode, minimum width=0.7cm, max width=0.7cm] {$+_{m_{6}}$};
\end{tikzpicture}

%% file: figures/sc2d_bitflip_generator.tex
\begin{tikzpicture}[very thick,scale=1]
\tikzstyle{tnnode}=[regular polygon,regular polygon sides=4,draw=black,fill=white,inner sep=0,outer sep=0]
\draw (1,1) -- (2,1);
\draw (1,1) -- (1,0);
\draw (1,1) -- (1,2);
\draw (1,3) -- (2,3);
\draw (1,3) -- (1,2);
\draw (1,3) -- (1,4);
\draw (3,1) -- (2,1);
\draw (3,1) -- (4,1);
\draw (3,1) -- (3,0);
\draw (3,1) -- (3,2);
\draw (3,3) -- (2,3);
\draw (3,3) -- (4,3);
\draw (3,3) -- (3,2);
\draw (3,3) -- (3,4);
\draw (5,1) -- (4,1);
\draw (5,1) -- (5,0);
\draw (5,1) -- (5,2);
\draw (5,3) -- (4,3);
\draw (5,3) -- (5,2);
\draw (5,3) -- (5,4);
\node (1_0) at (1,0) [tnnode,inner sep=-1pt] {$+_{p_{0}}$};
\node (1_2) at (1,2) [tnnode,inner sep=-1pt] {$+_{p_{1}}$};
\node (1_4) at (1,4) [tnnode,inner sep=-1pt] {$+_{p_{2}}$};
\node (3_0) at (3,0) [tnnode,inner sep=-1pt] {$+_{p_{5}}$};
\node (3_2) at (3,2) [tnnode,inner sep=-1pt] {$+_{p_{6}}$};
\node (3_4) at (3,4) [tnnode,inner sep=-1pt] {$+_{p_{7}}$};
\node (5_0) at (5,0) [tnnode,inner sep=-1pt] {$+_{p_{10}}$};
\node (5_2) at (5,2) [tnnode,inner sep=-1pt] {$+_{p_{11}}$};
\node (5_4) at (5,4) [tnnode,inner sep=-1pt] {$+_{p_{12}}$};
\node (2_1) at (2,1) [tnnode,inner sep=-1pt] {$+_{p_{3}}$};
\node (2_3) at (2,3) [tnnode,inner sep=-1pt] {$+_{p_{4}}$};
\node (4_1) at (4,1) [tnnode,inner sep=-1pt] {$+_{p_{8}}$};
\node (4_3) at (4,3) [tnnode,inner sep=-1pt] {$+_{p_{9}}$};
\node (1_1) at (1,1) [tnnode, minimum width=0.7cm, max width=0.7cm] {\eq};
\node (1_3) at (1,3) [tnnode, minimum width=0.7cm, max width=0.7cm] {\eq};
\node (3_1) at (3,1) [tnnode, minimum width=0.7cm, max width=0.7cm] {\eq};
\node (3_3) at (3,3) [tnnode, minimum width=0.7cm, max width=0.7cm] {\eq};
\node (5_1) at (5,1) [tnnode, minimum width=0.7cm, max width=0.7cm] {\eq};
\node (5_3) at (5,3) [tnnode, minimum width=0.7cm, max width=0.7cm] {\eq};
\end{tikzpicture}

%% file: figures/gate_splitting.tex
\begin{tikzpicture}[
square/.style={regular polygon,regular polygon sides=4,inner sep=0,minimum height=22, align=center}]
\draw[draw=none] (0,-3) rectangle ++(0,4);
\draw (-2,0) -- (2,0);
\draw[thick,densely dotted] (-2.1,0) -- (-2.27,0);
\draw[thick,densely dotted] (2.1,0) -- (2.27,0);
\draw (-1,-1) -- (-1,1);
\draw (1,-1) -- (1,1);
\draw (-1.7,-0.7) -- (-0.3,0.7);
\draw[thick,densely dotted] (-1.75,-0.75) -- (-1.87,-0.87);
\draw[thick,densely dotted] (-0.25,0.75) -- (-0.13,0.87);
\draw (0.3,-0.7) -- (1.7,0.7);
\draw[thick,densely dotted] (0.25,-0.75) -- (0.13,-0.87);
\draw[thick,densely dotted] (1.75,0.75) -- (1.87,0.87);
\node[square,draw,fill=white] at (-1,0) {A};
\node[square,draw,fill=white] at (1,0) {B};
\end{tikzpicture}
\begin{tikzpicture}
\draw[draw=none] (-0.6,-3) rectangle ++(1.2,4);
\node at (0,0){\large$=$};
\end{tikzpicture}
\begin{tikzpicture}[
square/.style={regular polygon,regular polygon sides=4,inner sep=-10,minimum height=22, align=center}]
\draw (-2,0) -- (-1,0);
\draw (1,0) -- (2,0);
\draw[thick,densely dotted] (-2.1,0) -- (-2.27,0);
\draw[thick,densely dotted] (2.1,0) -- (2.27,0);
\draw (-1,-3) -- (-1,1);
\draw (1,-3) -- (1,1);
\draw (-1.7,-0.7) -- (-0.3,0.7);
\draw (-1,-2) -- (1,-2);
\draw[thick,densely dotted] (-1.75,-0.75) -- (-1.87,-0.87);
\draw[thick,densely dotted] (-0.25,0.75) -- (-0.13,0.87);
\draw (0.3,-0.7) -- (1.7,0.7);
\draw[thick,densely dotted] (0.25,-0.75) -- (0.13,-0.87);
\draw[thick,densely dotted] (1.75,0.75) -- (1.87,0.87);
\node[square,draw,fill=white] at (-1,0) {$\mathrm{U_{\smash{A}}}$};
\node[square,draw,fill=white] at (1,0) {$\mathrm{U_{\smash{B}}}$};
\node[square,draw,fill=white] at (-1,-1) {$\mathrm{S_{\smash{A}}}$};
\node[square,draw,fill=white] at (1,-1) {$\mathrm{S_{\smash{B}}}$};
\node[square,draw,fill=white] at (-1,-2) {$\mathrm{V_{\smash{A}}}$};
\node[square,draw,fill=white] at (1,-2) {$\mathrm{V_{\smash{B}}}$};
\draw[densely dashed,red,thick,rounded corners=0.25cm] (-1.5,-2.5) rectangle ++(3.0,1.0);
\node at (2.8,-2) {\textbf{\color{red}2-qubit gate}};
\end{tikzpicture}

%% file: figures/tn_pointsec.tex
\tdplotsetmaincoords{75}{130} 
\begin{tikzpicture}[scale=2,tdplot_main_coords,thick]
\tikzstyle{tnnode}=[regular polygon,regular polygon sides=4,draw=black,fill=white,inner sep=0,outer sep=0]
\draw[color=black!40] (0,0,1) -- (0,0,0.5);
\draw[color=black!40] (0,0,1) -- (0,1,1);
\draw[color=black!40] (0,0,1) -- (0,0,2);
\draw[color=black!40] (0,0,2) -- (0,0,2.5);
\draw[color=black!40] (0,0,2) -- (0,1,2);
\draw[color=black!40] (0,0,2) -- (0,0,1);
\draw[color=black!40] (0,1,1) -- (0,1,0.5);
\draw[color=black!40] (0,1,1) -- (0,2,1);
\draw[color=black!40] (0,1,1) -- (0,0,1);
\draw[color=black!40] (0,1,1) -- (0,1,2);
\draw[color=black!40] (0,1,2) -- (0,1,2.5);
\draw[color=black!40] (0,1,2) -- (0,2,2);
\draw[color=black!40] (0,1,2) -- (0,0,2);
\draw[color=black!40] (0,1,2) -- (0,1,1);
\draw[color=black!40] (0,2,1) -- (0,2,0.5);
\draw[color=black!40] (0,2,1) -- (0,1,1);
\draw[color=black!40] (0,2,1) -- (0,2,2);
\draw[color=black!40] (0,2,2) -- (0,2,2.5);
\draw[color=black!40] (0,2,2) -- (0,1,2);
\draw[color=black!40] (0,2,2) -- (0,2,1);
\draw[color=black!55] (0,0,1) -- (0+1,0,1);
\draw[color=black!55] (0,0,2) -- (0+1,0,2);
\draw[color=black!55] (0,1,1) -- (0+1,1,1);
\draw[color=black!55] (0,1,2) -- (0+1,1,2);
\draw[color=black!55] (0,2,1) -- (0+1,2,1);
\draw[color=black!55] (0,2,2) -- (0+1,2,2);
\draw (0,0,1) node[tnnode,draw=black!40,text=black!40] {$+$};
\draw (0,0,0.5) node[tnnode,draw=black!40,text=black!40] {\eq};
\draw (0,0.5,1.0) node[tnnode,draw=black!40,text=black!40] {\eq};
\draw (0,0.0,1.5) node[tnnode,draw=black!40,text=black!40] {\eq};
\draw (0,0,2) node[tnnode,draw=black!40,text=black!40] {$+$};
\draw (0,0,2.5) node[tnnode,draw=black!40,text=black!40] {\eq};
\draw (0,0.5,2.0) node[tnnode,draw=black!40,text=black!40] {\eq};
\draw (0,0.0,1.5) node[tnnode,draw=black!40,text=black!40] {\eq};
\draw (0,1,1) node[tnnode,draw=black!40,text=black!40] {$+$};
\draw (0,1,0.5) node[tnnode,draw=black!40,text=black!40] {\eq};
\draw (0,1.5,1.0) node[tnnode,draw=black!40,text=black!40] {\eq};
\draw (0,0.5,1.0) node[tnnode,draw=black!40,text=black!40] {\eq};
\draw (0,1.0,1.5) node[tnnode,draw=black!40,text=black!40] {\eq};
\draw (0,1,2) node[tnnode,draw=black!40,text=black!40] {$+$};
\draw (0,1,2.5) node[tnnode,draw=black!40,text=black!40] {\eq};
\draw (0,1.5,2.0) node[tnnode,draw=black!40,text=black!40] {\eq};
\draw (0,0.5,2.0) node[tnnode,draw=black!40,text=black!40] {\eq};
\draw (0,1.0,1.5) node[tnnode,draw=black!40,text=black!40] {\eq};
\draw (0,2,1) node[tnnode,draw=black!40,text=black!40] {$+$};
\draw (0,2,0.5) node[tnnode,draw=black!40,text=black!40] {\eq};
\draw (0,1.5,1.0) node[tnnode,draw=black!40,text=black!40] {\eq};
\draw (0,2.0,1.5) node[tnnode,draw=black!40,text=black!40] {\eq};
\draw (0,2,2) node[tnnode,draw=black!40,text=black!40] {$+$};
\draw (0,2,2.5) node[tnnode,draw=black!40,text=black!40] {\eq};
\draw (0,1.5,2.0) node[tnnode,draw=black!40,text=black!40] {\eq};
\draw (0,2.0,1.5) node[tnnode,draw=black!40,text=black!40] {\eq};
\draw (0.5,0,1) node[tnnode,draw=black!55,text=black!55] {\eq};
\draw (0.5,0,2) node[tnnode,draw=black!55,text=black!55] {\eq};
\draw (0.5,1,1) node[tnnode,draw=black!55,text=black!55] {\eq};
\draw (0.5,1,2) node[tnnode,draw=black!55,text=black!55] {\eq};
\draw (0.5,2,1) node[tnnode,draw=black!55,text=black!55] {\eq};
\draw (0.5,2,2) node[tnnode,draw=black!55,text=black!55] {\eq};
\draw[color=black!70] (1,0,1) -- (1,0,0.5);
\draw[color=black!70] (1,0,1) -- (1,1,1);
\draw[color=black!70] (1,0,1) -- (1,0,2);
\draw[color=black!70] (1,0,2) -- (1,0,2.5);
\draw[color=black!70] (1,0,2) -- (1,1,2);
\draw[color=black!70] (1,0,2) -- (1,0,1);
\draw[color=black!70] (1,1,1) -- (1,1,0.5);
\draw[color=black!70] (1,1,1) -- (1,2,1);
\draw[color=black!70] (1,1,1) -- (1,0,1);
\draw[color=black!70] (1,1,1) -- (1,1,2);
\draw[color=black!70] (1,1,2) -- (1,1,2.5);
\draw[color=black!70] (1,1,2) -- (1,2,2);
\draw[color=black!70] (1,1,2) -- (1,0,2);
\draw[color=black!70] (1,1,2) -- (1,1,1);
\draw[color=black!70] (1,2,1) -- (1,2,0.5);
\draw[color=black!70] (1,2,1) -- (1,1,1);
\draw[color=black!70] (1,2,1) -- (1,2,2);
\draw[color=black!70] (1,2,2) -- (1,2,2.5);
\draw[color=black!70] (1,2,2) -- (1,1,2);
\draw[color=black!70] (1,2,2) -- (1,2,1);
\draw[color=black!85] (1,0,1) -- (1+1,0,1);
\draw[color=black!85] (1,0,2) -- (1+1,0,2);
\draw[color=black!85] (1,1,1) -- (1+1,1,1);
\draw[color=black!85] (1,1,2) -- (1+1,1,2);
\draw[color=black!85] (1,2,1) -- (1+1,2,1);
\draw[color=black!85] (1,2,2) -- (1+1,2,2);
\draw (1,0,1) node[tnnode,draw=black!70,text=black!70] {$+$};
\draw (1,0,0.5) node[tnnode,draw=black!70,text=black!70] {\eq};
\draw (1,0.5,1.0) node[tnnode,draw=black!70,text=black!70] {\eq};
\draw (1,0.0,1.5) node[tnnode,draw=black!70,text=black!70] {\eq};
\draw (1,0,2) node[tnnode,draw=black!70,text=black!70] {$+$};
\draw (1,0,2.5) node[tnnode,draw=black!70,text=black!70] {\eq};
\draw (1,0.5,2.0) node[tnnode,draw=black!70,text=black!70] {\eq};
\draw (1,0.0,1.5) node[tnnode,draw=black!70,text=black!70] {\eq};
\draw (1,1,1) node[tnnode,draw=black!70,text=black!70] {$+$};
\draw (1,1,0.5) node[tnnode,draw=black!70,text=black!70] {\eq};
\draw (1,1.5,1.0) node[tnnode,draw=black!70,text=black!70] {\eq};
\draw (1,0.5,1.0) node[tnnode,draw=black!70,text=black!70] {\eq};
\draw (1,1.0,1.5) node[tnnode,draw=black!70,text=black!70] {\eq};
\draw (1,1,2) node[tnnode,draw=black!70,text=black!70] {$+$};
\draw (1,1,2.5) node[tnnode,draw=black!70,text=black!70] {\eq};
\draw (1,1.5,2.0) node[tnnode,draw=black!70,text=black!70] {\eq};
\draw (1,0.5,2.0) node[tnnode,draw=black!70,text=black!70] {\eq};
\draw (1,1.0,1.5) node[tnnode,draw=black!70,text=black!70] {\eq};
\draw (1,2,1) node[tnnode,draw=black!70,text=black!70] {$+$};
\draw (1,2,0.5) node[tnnode,draw=black!70,text=black!70] {\eq};
\draw (1,1.5,1.0) node[tnnode,draw=black!70,text=black!70] {\eq};
\draw (1,2.0,1.5) node[tnnode,draw=black!70,text=black!70] {\eq};
\draw (1,2,2) node[tnnode,draw=black!70,text=black!70] {$+$};
\draw (1,2,2.5) node[tnnode,draw=black!70,text=black!70] {\eq};
\draw (1,1.5,2.0) node[tnnode,draw=black!70,text=black!70] {\eq};
\draw (1,2.0,1.5) node[tnnode,draw=black!70,text=black!70] {\eq};
\draw (1.5,0,1) node[tnnode,draw=black!85,text=black!85] {\eq};
\draw (1.5,0,2) node[tnnode,draw=black!85,text=black!85] {\eq};
\draw (1.5,1,1) node[tnnode,draw=black!85,text=black!85] {\eq};
\draw (1.5,1,2) node[tnnode,draw=black!85,text=black!85] {\eq};
\draw (1.5,2,1) node[tnnode,draw=black!85,text=black!85] {\eq};
\draw (1.5,2,2) node[tnnode,draw=black!85,text=black!85] {\eq};
\draw[color=black!100] (2,0,1) -- (2,0,0.5);
\draw[color=black!100] (2,0,1) -- (2,1,1);
\draw[color=black!100] (2,0,1) -- (2,0,2);
\draw[color=black!100] (2,0,2) -- (2,0,2.5);
\draw[color=black!100] (2,0,2) -- (2,1,2);
\draw[color=black!100] (2,0,2) -- (2,0,1);
\draw[color=black!100] (2,1,1) -- (2,1,0.5);
\draw[color=black!100] (2,1,1) -- (2,2,1);
\draw[color=black!100] (2,1,1) -- (2,0,1);
\draw[color=black!100] (2,1,1) -- (2,1,2);
\draw[color=black!100] (2,1,2) -- (2,1,2.5);
\draw[color=black!100] (2,1,2) -- (2,2,2);
\draw[color=black!100] (2,1,2) -- (2,0,2);
\draw[color=black!100] (2,1,2) -- (2,1,1);
\draw[color=black!100] (2,2,1) -- (2,2,0.5);
\draw[color=black!100] (2,2,1) -- (2,1,1);
\draw[color=black!100] (2,2,1) -- (2,2,2);
\draw[color=black!100] (2,2,2) -- (2,2,2.5);
\draw[color=black!100] (2,2,2) -- (2,1,2);
\draw[color=black!100] (2,2,2) -- (2,2,1);
\draw (2,0,1) node[tnnode,draw=black!100,text=black!100] {$+$};
\draw (2,0,0.5) node[tnnode,draw=black!100,text=black!100] {\eq};
\draw (2,0.5,1.0) node[tnnode,draw=black!100,text=black!100] {\eq};
\draw (2,0.0,1.5) node[tnnode,draw=black!100,text=black!100] {\eq};
\draw (2,0,2) node[tnnode,draw=black!100,text=black!100] {$+$};
\draw (2,0,2.5) node[tnnode,draw=black!100,text=black!100] {\eq};
\draw (2,0.5,2.0) node[tnnode,draw=black!100,text=black!100] {\eq};
\draw (2,0.0,1.5) node[tnnode,draw=black!100,text=black!100] {\eq};
\draw (2,1,1) node[tnnode,draw=black!100,text=black!100] {$+$};
\draw (2,1,0.5) node[tnnode,draw=black!100,text=black!100] {\eq};
\draw (2,1.5,1.0) node[tnnode,draw=black!100,text=black!100] {\eq};
\draw (2,0.5,1.0) node[tnnode,draw=black!100,text=black!100] {\eq};
\draw (2,1.0,1.5) node[tnnode,draw=black!100,text=black!100] {\eq};
\draw (2,1,2) node[tnnode,draw=black!100,text=black!100] {$+$};
\draw (2,1,2.5) node[tnnode,draw=black!100,text=black!100] {\eq};
\draw (2,1.5,2.0) node[tnnode,draw=black!100,text=black!100] {\eq};
\draw (2,0.5,2.0) node[tnnode,draw=black!100,text=black!100] {\eq};
\draw (2,1.0,1.5) node[tnnode,draw=black!100,text=black!100] {\eq};
\draw (2,2,1) node[tnnode,draw=black!100,text=black!100] {$+$};
\draw (2,2,0.5) node[tnnode,draw=black!100,text=black!100] {\eq};
\draw (2,1.5,1.0) node[tnnode,draw=black!100,text=black!100] {\eq};
\draw (2,2.0,1.5) node[tnnode,draw=black!100,text=black!100] {\eq};
\draw (2,2,2) node[tnnode,draw=black!100,text=black!100] {$+$};
\draw (2,2,2.5) node[tnnode,draw=black!100,text=black!100] {\eq};
\draw (2,1.5,2.0) node[tnnode,draw=black!100,text=black!100] {\eq};
\draw (2,2.0,1.5) node[tnnode,draw=black!100,text=black!100] {\eq};
\end{tikzpicture}

%% file: figures/tn_depol.tex
\tdplotsetmaincoords{75}{130} 
\begin{tikzpicture}[scale=1.5,tdplot_main_coords,thick]
\tikzstyle{tnnode}=[regular polygon,regular polygon sides=4,draw=black,fill=white,inner sep=0,outer sep=0]
\draw[color=black!40] (0,0,1) -- (0,1,1);
\draw[color=black!40] (0,0,1) -- (0,0,2);
\draw[color=black!40] (0,0,2) -- (0,1,2);
\draw[color=black!40] (0,0,2) -- (0,0,3);
\draw[color=black!40] (0,0,2) -- (0,0,1);
\draw[color=black!40] (0,0,3) -- (0,1,3);
\draw[color=black!40] (0,0,3) -- (0,0,4);
\draw[color=black!40] (0,0,3) -- (0,0,2);
\draw[color=black!40] (0,0,4) -- (0,1,4);
\draw[color=black!40] (0,0,4) -- (0,0,5);
\draw[color=black!40] (0,0,4) -- (0,0,3);
\draw[color=black!40] (0,0,5) -- (0,1,5);
\draw[color=black!40] (0,0,5) -- (0,0,4);
\draw[color=black!40] (0,1,1) -- (0,2,1);
\draw[color=black!40] (0,1,1) -- (0,0,1);
\draw[color=black!40] (0,1,1) -- (0,1,2);
\draw[color=black!40] (0,1,2) -- (0,2,2);
\draw[color=black!40] (0,1,2) -- (0,0,2);
\draw[color=black!40] (0,1,2) -- (0,1,3);
\draw[color=black!40] (0,1,2) -- (0,1,1);
\draw[color=black!40] (0,1,3) -- (0,2,3);
\draw[color=black!40] (0,1,3) -- (0,0,3);
\draw[color=black!40] (0,1,3) -- (0,1,4);
\draw[color=black!40] (0,1,3) -- (0,1,2);
\draw[color=black!40] (0,1,4) -- (0,2,4);
\draw[color=black!40] (0,1,4) -- (0,0,4);
\draw[color=black!40] (0,1,4) -- (0,1,5);
\draw[color=black!40] (0,1,4) -- (0,1,3);
\draw[color=black!40] (0,1,5) -- (0,2,5);
\draw[color=black!40] (0,1,5) -- (0,0,5);
\draw[color=black!40] (0,1,5) -- (0,1,4);
\draw[color=black!40] (0,2,1) -- (0,3,1);
\draw[color=black!40] (0,2,1) -- (0,1,1);
\draw[color=black!40] (0,2,1) -- (0,2,2);
\draw[color=black!40] (0,2,2) -- (0,3,2);
\draw[color=black!40] (0,2,2) -- (0,1,2);
\draw[color=black!40] (0,2,2) -- (0,2,3);
\draw[color=black!40] (0,2,2) -- (0,2,1);
\draw[color=black!40] (0,2,3) -- (0,3,3);
\draw[color=black!40] (0,2,3) -- (0,1,3);
\draw[color=black!40] (0,2,3) -- (0,2,4);
\draw[color=black!40] (0,2,3) -- (0,2,2);
\draw[color=black!40] (0,2,4) -- (0,3,4);
\draw[color=black!40] (0,2,4) -- (0,1,4);
\draw[color=black!40] (0,2,4) -- (0,2,5);
\draw[color=black!40] (0,2,4) -- (0,2,3);
\draw[color=black!40] (0,2,5) -- (0,3,5);
\draw[color=black!40] (0,2,5) -- (0,1,5);
\draw[color=black!40] (0,2,5) -- (0,2,4);
\draw[color=black!40] (0,3,1) -- (0,4,1);
\draw[color=black!40] (0,3,1) -- (0,2,1);
\draw[color=black!40] (0,3,1) -- (0,3,2);
\draw[color=black!40] (0,3,2) -- (0,4,2);
\draw[color=black!40] (0,3,2) -- (0,2,2);
\draw[color=black!40] (0,3,2) -- (0,3,3);
\draw[color=black!40] (0,3,2) -- (0,3,1);
\draw[color=black!40] (0,3,3) -- (0,4,3);
\draw[color=black!40] (0,3,3) -- (0,2,3);
\draw[color=black!40] (0,3,3) -- (0,3,4);
\draw[color=black!40] (0,3,3) -- (0,3,2);
\draw[color=black!40] (0,3,4) -- (0,4,4);
\draw[color=black!40] (0,3,4) -- (0,2,4);
\draw[color=black!40] (0,3,4) -- (0,3,5);
\draw[color=black!40] (0,3,4) -- (0,3,3);
\draw[color=black!40] (0,3,5) -- (0,4,5);
\draw[color=black!40] (0,3,5) -- (0,2,5);
\draw[color=black!40] (0,3,5) -- (0,3,4);
\draw[color=black!40] (0,4,1) -- (0,3,1);
\draw[color=black!40] (0,4,1) -- (0,4,2);
\draw[color=black!40] (0,4,2) -- (0,3,2);
\draw[color=black!40] (0,4,2) -- (0,4,3);
\draw[color=black!40] (0,4,2) -- (0,4,1);
\draw[color=black!40] (0,4,3) -- (0,3,3);
\draw[color=black!40] (0,4,3) -- (0,4,4);
\draw[color=black!40] (0,4,3) -- (0,4,2);
\draw[color=black!40] (0,4,4) -- (0,3,4);
\draw[color=black!40] (0,4,4) -- (0,4,5);
\draw[color=black!40] (0,4,4) -- (0,4,3);
\draw[color=black!40] (0,4,5) -- (0,3,5);
\draw[color=black!40] (0,4,5) -- (0,4,4);
\draw[color=black!48] (0,0,1) -- (0+1,0,1);
\draw[color=black!48] (0,0,2) -- (0+1,0,2);
\draw[color=black!48] (0,0,3) -- (0+1,0,3);
\draw[color=black!48] (0,0,4) -- (0+1,0,4);
\draw[color=black!48] (0,0,5) -- (0+1,0,5);
\draw[color=black!48] (0,1,2) -- (0+1,1,2);
\draw[color=black!48] (0,1,4) -- (0+1,1,4);
\draw[color=black!48] (0,2,2) -- (0+1,2,2);
\draw[color=black!48] (0,2,4) -- (0+1,2,4);
\draw[color=black!48] (0,3,2) -- (0+1,3,2);
\draw[color=black!48] (0,3,4) -- (0+1,3,4);
\draw[color=black!48] (0,4,2) -- (0+1,4,2);
\draw[color=black!48] (0,4,4) -- (0+1,4,4);
\draw (0,0,1) node[tnnode,draw=black!40,text=black!40] {$d$};
\draw (0,0,2) node[tnnode,draw=black!40,text=black!40] {$+$};
\draw (0,0,3) node[tnnode,draw=black!40,text=black!40] {$d$};
\draw (0,0,4) node[tnnode,draw=black!40,text=black!40] {$+$};
\draw (0,0,5) node[tnnode,draw=black!40,text=black!40] {$d$};
\draw (0,1,1) node[tnnode,draw=black!40,text=black!40] {$+$};
\draw (0,1,2) node[tnnode,draw=black!40,text=black!40] {$d$};
\draw (0,1,3) node[tnnode,draw=black!40,text=black!40] {$+$};
\draw (0,1,4) node[tnnode,draw=black!40,text=black!40] {$d$};
\draw (0,1,5) node[tnnode,draw=black!40,text=black!40] {$+$};
\draw (0,2,1) node[tnnode,draw=black!40,text=black!40] {$d$};
\draw (0,2,2) node[tnnode,draw=black!40,text=black!40] {$+$};
\draw (0,2,3) node[tnnode,draw=black!40,text=black!40] {$d$};
\draw (0,2,4) node[tnnode,draw=black!40,text=black!40] {$+$};
\draw (0,2,5) node[tnnode,draw=black!40,text=black!40] {$d$};
\draw (0,3,1) node[tnnode,draw=black!40,text=black!40] {$+$};
\draw (0,3,2) node[tnnode,draw=black!40,text=black!40] {$d$};
\draw (0,3,3) node[tnnode,draw=black!40,text=black!40] {$+$};
\draw (0,3,4) node[tnnode,draw=black!40,text=black!40] {$d$};
\draw (0,3,5) node[tnnode,draw=black!40,text=black!40] {$+$};
\draw (0,4,1) node[tnnode,draw=black!40,text=black!40] {$d$};
\draw (0,4,2) node[tnnode,draw=black!40,text=black!40] {$+$};
\draw (0,4,3) node[tnnode,draw=black!40,text=black!40] {$d$};
\draw (0,4,4) node[tnnode,draw=black!40,text=black!40] {$+$};
\draw (0,4,5) node[tnnode,draw=black!40,text=black!40] {$d$};
\draw[color=black!55] (1,0,1) -- (1,0,2);
\draw[color=black!55] (1,0,2) -- (1,1,2);
\draw[color=black!55] (1,0,2) -- (1,0,3);
\draw[color=black!55] (1,0,2) -- (1,0,1);
\draw[color=black!55] (1,0,3) -- (1,0,4);
\draw[color=black!55] (1,0,3) -- (1,0,2);
\draw[color=black!55] (1,0,4) -- (1,1,4);
\draw[color=black!55] (1,0,4) -- (1,0,5);
\draw[color=black!55] (1,0,4) -- (1,0,3);
\draw[color=black!55] (1,0,5) -- (1,0,4);
\draw[color=black!55] (1,1,2) -- (1,2,2);
\draw[color=black!55] (1,1,2) -- (1,0,2);
\draw[color=black!55] (1,1,4) -- (1,2,4);
\draw[color=black!55] (1,1,4) -- (1,0,4);
\draw[color=black!55] (1,2,2) -- (1,3,2);
\draw[color=black!55] (1,2,2) -- (1,1,2);
\draw[color=black!55] (1,2,4) -- (1,3,4);
\draw[color=black!55] (1,2,4) -- (1,1,4);
\draw[color=black!55] (1,3,2) -- (1,4,2);
\draw[color=black!55] (1,3,2) -- (1,2,2);
\draw[color=black!55] (1,3,4) -- (1,4,4);
\draw[color=black!55] (1,3,4) -- (1,2,4);
\draw[color=black!55] (1,4,2) -- (1,3,2);
\draw[color=black!55] (1,4,4) -- (1,3,4);
\draw[color=black!62] (1,0,1) -- (1+1,0,1);
\draw[color=black!62] (1,0,2) -- (1+1,0,2);
\draw[color=black!62] (1,0,3) -- (1+1,0,3);
\draw[color=black!62] (1,0,4) -- (1+1,0,4);
\draw[color=black!62] (1,0,5) -- (1+1,0,5);
\draw[color=black!62] (1,1,2) -- (1+1,1,2);
\draw[color=black!62] (1,1,4) -- (1+1,1,4);
\draw[color=black!62] (1,2,2) -- (1+1,2,2);
\draw[color=black!62] (1,2,4) -- (1+1,2,4);
\draw[color=black!62] (1,3,2) -- (1+1,3,2);
\draw[color=black!62] (1,3,4) -- (1+1,3,4);
\draw[color=black!62] (1,4,2) -- (1+1,4,2);
\draw[color=black!62] (1,4,4) -- (1+1,4,4);
\draw (1,0,1) node[tnnode,draw=black!55,text=black!55] {$+$};
\draw (1,0,2) node[tnnode,draw=black!55,text=black!55] {$d$};
\draw (1,0,3) node[tnnode,draw=black!55,text=black!55] {$+$};
\draw (1,0,4) node[tnnode,draw=black!55,text=black!55] {$d$};
\draw (1,0,5) node[tnnode,draw=black!55,text=black!55] {$+$};
\draw (1,1,2) node[tnnode,draw=black!55,text=black!55] {$+$};
\draw (1,1,4) node[tnnode,draw=black!55,text=black!55] {$+$};
\draw (1,2,2) node[tnnode,draw=black!55,text=black!55] {$d$};
\draw (1,2,4) node[tnnode,draw=black!55,text=black!55] {$d$};
\draw (1,3,2) node[tnnode,draw=black!55,text=black!55] {$+$};
\draw (1,3,4) node[tnnode,draw=black!55,text=black!55] {$+$};
\draw (1,4,2) node[tnnode,draw=black!55,text=black!55] {$d$};
\draw (1,4,4) node[tnnode,draw=black!55,text=black!55] {$d$};
\draw[color=black!70] (2,0,1) -- (2,1,1);
\draw[color=black!70] (2,0,1) -- (2,0,2);
\draw[color=black!70] (2,0,2) -- (2,1,2);
\draw[color=black!70] (2,0,2) -- (2,0,3);
\draw[color=black!70] (2,0,2) -- (2,0,1);
\draw[color=black!70] (2,0,3) -- (2,1,3);
\draw[color=black!70] (2,0,3) -- (2,0,4);
\draw[color=black!70] (2,0,3) -- (2,0,2);
\draw[color=black!70] (2,0,4) -- (2,1,4);
\draw[color=black!70] (2,0,4) -- (2,0,5);
\draw[color=black!70] (2,0,4) -- (2,0,3);
\draw[color=black!70] (2,0,5) -- (2,1,5);
\draw[color=black!70] (2,0,5) -- (2,0,4);
\draw[color=black!70] (2,1,1) -- (2,2,1);
\draw[color=black!70] (2,1,1) -- (2,0,1);
\draw[color=black!70] (2,1,1) -- (2,1,2);
\draw[color=black!70] (2,1,2) -- (2,2,2);
\draw[color=black!70] (2,1,2) -- (2,0,2);
\draw[color=black!70] (2,1,2) -- (2,1,3);
\draw[color=black!70] (2,1,2) -- (2,1,1);
\draw[color=black!70] (2,1,3) -- (2,2,3);
\draw[color=black!70] (2,1,3) -- (2,0,3);
\draw[color=black!70] (2,1,3) -- (2,1,4);
\draw[color=black!70] (2,1,3) -- (2,1,2);
\draw[color=black!70] (2,1,4) -- (2,2,4);
\draw[color=black!70] (2,1,4) -- (2,0,4);
\draw[color=black!70] (2,1,4) -- (2,1,5);
\draw[color=black!70] (2,1,4) -- (2,1,3);
\draw[color=black!70] (2,1,5) -- (2,2,5);
\draw[color=black!70] (2,1,5) -- (2,0,5);
\draw[color=black!70] (2,1,5) -- (2,1,4);
\draw[color=black!70] (2,2,1) -- (2,3,1);
\draw[color=black!70] (2,2,1) -- (2,1,1);
\draw[color=black!70] (2,2,1) -- (2,2,2);
\draw[color=black!70] (2,2,2) -- (2,3,2);
\draw[color=black!70] (2,2,2) -- (2,1,2);
\draw[color=black!70] (2,2,2) -- (2,2,3);
\draw[color=black!70] (2,2,2) -- (2,2,1);
\draw[color=black!70] (2,2,3) -- (2,3,3);
\draw[color=black!70] (2,2,3) -- (2,1,3);
\draw[color=black!70] (2,2,3) -- (2,2,4);
\draw[color=black!70] (2,2,3) -- (2,2,2);
\draw[color=black!70] (2,2,4) -- (2,3,4);
\draw[color=black!70] (2,2,4) -- (2,1,4);
\draw[color=black!70] (2,2,4) -- (2,2,5);
\draw[color=black!70] (2,2,4) -- (2,2,3);
\draw[color=black!70] (2,2,5) -- (2,3,5);
\draw[color=black!70] (2,2,5) -- (2,1,5);
\draw[color=black!70] (2,2,5) -- (2,2,4);
\draw[color=black!70] (2,3,1) -- (2,4,1);
\draw[color=black!70] (2,3,1) -- (2,2,1);
\draw[color=black!70] (2,3,1) -- (2,3,2);
\draw[color=black!70] (2,3,2) -- (2,4,2);
\draw[color=black!70] (2,3,2) -- (2,2,2);
\draw[color=black!70] (2,3,2) -- (2,3,3);
\draw[color=black!70] (2,3,2) -- (2,3,1);
\draw[color=black!70] (2,3,3) -- (2,4,3);
\draw[color=black!70] (2,3,3) -- (2,2,3);
\draw[color=black!70] (2,3,3) -- (2,3,4);
\draw[color=black!70] (2,3,3) -- (2,3,2);
\draw[color=black!70] (2,3,4) -- (2,4,4);
\draw[color=black!70] (2,3,4) -- (2,2,4);
\draw[color=black!70] (2,3,4) -- (2,3,5);
\draw[color=black!70] (2,3,4) -- (2,3,3);
\draw[color=black!70] (2,3,5) -- (2,4,5);
\draw[color=black!70] (2,3,5) -- (2,2,5);
\draw[color=black!70] (2,3,5) -- (2,3,4);
\draw[color=black!70] (2,4,1) -- (2,3,1);
\draw[color=black!70] (2,4,1) -- (2,4,2);
\draw[color=black!70] (2,4,2) -- (2,3,2);
\draw[color=black!70] (2,4,2) -- (2,4,3);
\draw[color=black!70] (2,4,2) -- (2,4,1);
\draw[color=black!70] (2,4,3) -- (2,3,3);
\draw[color=black!70] (2,4,3) -- (2,4,4);
\draw[color=black!70] (2,4,3) -- (2,4,2);
\draw[color=black!70] (2,4,4) -- (2,3,4);
\draw[color=black!70] (2,4,4) -- (2,4,5);
\draw[color=black!70] (2,4,4) -- (2,4,3);
\draw[color=black!70] (2,4,5) -- (2,3,5);
\draw[color=black!70] (2,4,5) -- (2,4,4);
\draw[color=black!78] (2,0,1) -- (2+1,0,1);
\draw[color=black!78] (2,0,2) -- (2+1,0,2);
\draw[color=black!78] (2,0,3) -- (2+1,0,3);
\draw[color=black!78] (2,0,4) -- (2+1,0,4);
\draw[color=black!78] (2,0,5) -- (2+1,0,5);
\draw[color=black!78] (2,1,2) -- (2+1,1,2);
\draw[color=black!78] (2,1,4) -- (2+1,1,4);
\draw[color=black!78] (2,2,2) -- (2+1,2,2);
\draw[color=black!78] (2,2,4) -- (2+1,2,4);
\draw[color=black!78] (2,3,2) -- (2+1,3,2);
\draw[color=black!78] (2,3,4) -- (2+1,3,4);
\draw[color=black!78] (2,4,2) -- (2+1,4,2);
\draw[color=black!78] (2,4,4) -- (2+1,4,4);
\draw (2,0,1) node[tnnode,draw=black!70,text=black!70] {$d$};
\draw (2,0,2) node[tnnode,draw=black!70,text=black!70] {$+$};
\draw (2,0,3) node[tnnode,draw=black!70,text=black!70] {$d$};
\draw (2,0,4) node[tnnode,draw=black!70,text=black!70] {$+$};
\draw (2,0,5) node[tnnode,draw=black!70,text=black!70] {$d$};
\draw (2,1,1) node[tnnode,draw=black!70,text=black!70] {$+$};
\draw (2,1,2) node[tnnode,draw=black!70,text=black!70] {$d$};
\draw (2,1,3) node[tnnode,draw=black!70,text=black!70] {$+$};
\draw (2,1,4) node[tnnode,draw=black!70,text=black!70] {$d$};
\draw (2,1,5) node[tnnode,draw=black!70,text=black!70] {$+$};
\draw (2,2,1) node[tnnode,draw=black!70,text=black!70] {$d$};
\draw (2,2,2) node[tnnode,draw=black!70,text=black!70] {$+$};
\draw (2,2,3) node[tnnode,draw=black!70,text=black!70] {$d$};
\draw (2,2,4) node[tnnode,draw=black!70,text=black!70] {$+$};
\draw (2,2,5) node[tnnode,draw=black!70,text=black!70] {$d$};
\draw (2,3,1) node[tnnode,draw=black!70,text=black!70] {$+$};
\draw (2,3,2) node[tnnode,draw=black!70,text=black!70] {$d$};
\draw (2,3,3) node[tnnode,draw=black!70,text=black!70] {$+$};
\draw (2,3,4) node[tnnode,draw=black!70,text=black!70] {$d$};
\draw (2,3,5) node[tnnode,draw=black!70,text=black!70] {$+$};
\draw (2,4,1) node[tnnode,draw=black!70,text=black!70] {$d$};
\draw (2,4,2) node[tnnode,draw=black!70,text=black!70] {$+$};
\draw (2,4,3) node[tnnode,draw=black!70,text=black!70] {$d$};
\draw (2,4,4) node[tnnode,draw=black!70,text=black!70] {$+$};
\draw (2,4,5) node[tnnode,draw=black!70,text=black!70] {$d$};
\draw[color=black!85] (3,0,1) -- (3,0,2);
\draw[color=black!85] (3,0,2) -- (3,1,2);
\draw[color=black!85] (3,0,2) -- (3,0,3);
\draw[color=black!85] (3,0,2) -- (3,0,1);
\draw[color=black!85] (3,0,3) -- (3,0,4);
\draw[color=black!85] (3,0,3) -- (3,0,2);
\draw[color=black!85] (3,0,4) -- (3,1,4);
\draw[color=black!85] (3,0,4) -- (3,0,5);
\draw[color=black!85] (3,0,4) -- (3,0,3);
\draw[color=black!85] (3,0,5) -- (3,0,4);
\draw[color=black!85] (3,1,2) -- (3,2,2);
\draw[color=black!85] (3,1,2) -- (3,0,2);
\draw[color=black!85] (3,1,4) -- (3,2,4);
\draw[color=black!85] (3,1,4) -- (3,0,4);
\draw[color=black!85] (3,2,2) -- (3,3,2);
\draw[color=black!85] (3,2,2) -- (3,1,2);
\draw[color=black!85] (3,2,4) -- (3,3,4);
\draw[color=black!85] (3,2,4) -- (3,1,4);
\draw[color=black!85] (3,3,2) -- (3,4,2);
\draw[color=black!85] (3,3,2) -- (3,2,2);
\draw[color=black!85] (3,3,4) -- (3,4,4);
\draw[color=black!85] (3,3,4) -- (3,2,4);
\draw[color=black!85] (3,4,2) -- (3,3,2);
\draw[color=black!85] (3,4,4) -- (3,3,4);
\draw[color=black!92] (3,0,1) -- (3+1,0,1);
\draw[color=black!92] (3,0,2) -- (3+1,0,2);
\draw[color=black!92] (3,0,3) -- (3+1,0,3);
\draw[color=black!92] (3,0,4) -- (3+1,0,4);
\draw[color=black!92] (3,0,5) -- (3+1,0,5);
\draw[color=black!92] (3,1,2) -- (3+1,1,2);
\draw[color=black!92] (3,1,4) -- (3+1,1,4);
\draw[color=black!92] (3,2,2) -- (3+1,2,2);
\draw[color=black!92] (3,2,4) -- (3+1,2,4);
\draw[color=black!92] (3,3,2) -- (3+1,3,2);
\draw[color=black!92] (3,3,4) -- (3+1,3,4);
\draw[color=black!92] (3,4,2) -- (3+1,4,2);
\draw[color=black!92] (3,4,4) -- (3+1,4,4);
\draw (3,0,1) node[tnnode,draw=black!85,text=black!85] {$+$};
\draw (3,0,2) node[tnnode,draw=black!85,text=black!85] {$d$};
\draw (3,0,3) node[tnnode,draw=black!85,text=black!85] {$+$};
\draw (3,0,4) node[tnnode,draw=black!85,text=black!85] {$d$};
\draw (3,0,5) node[tnnode,draw=black!85,text=black!85] {$+$};
\draw (3,1,2) node[tnnode,draw=black!85,text=black!85] {$+$};
\draw (3,1,4) node[tnnode,draw=black!85,text=black!85] {$+$};
\draw (3,2,2) node[tnnode,draw=black!85,text=black!85] {$d$};
\draw (3,2,4) node[tnnode,draw=black!85,text=black!85] {$d$};
\draw (3,3,2) node[tnnode,draw=black!85,text=black!85] {$+$};
\draw (3,3,4) node[tnnode,draw=black!85,text=black!85] {$+$};
\draw (3,4,2) node[tnnode,draw=black!85,text=black!85] {$d$};
\draw (3,4,4) node[tnnode,draw=black!85,text=black!85] {$d$};
\draw[color=black!100] (4,0,1) -- (4,1,1);
\draw[color=black!100] (4,0,1) -- (4,0,2);
\draw[color=black!100] (4,0,2) -- (4,1,2);
\draw[color=black!100] (4,0,2) -- (4,0,3);
\draw[color=black!100] (4,0,2) -- (4,0,1);
\draw[color=black!100] (4,0,3) -- (4,1,3);
\draw[color=black!100] (4,0,3) -- (4,0,4);
\draw[color=black!100] (4,0,3) -- (4,0,2);
\draw[color=black!100] (4,0,4) -- (4,1,4);
\draw[color=black!100] (4,0,4) -- (4,0,5);
\draw[color=black!100] (4,0,4) -- (4,0,3);
\draw[color=black!100] (4,0,5) -- (4,1,5);
\draw[color=black!100] (4,0,5) -- (4,0,4);
\draw[color=black!100] (4,1,1) -- (4,2,1);
\draw[color=black!100] (4,1,1) -- (4,0,1);
\draw[color=black!100] (4,1,1) -- (4,1,2);
\draw[color=black!100] (4,1,2) -- (4,2,2);
\draw[color=black!100] (4,1,2) -- (4,0,2);
\draw[color=black!100] (4,1,2) -- (4,1,3);
\draw[color=black!100] (4,1,2) -- (4,1,1);
\draw[color=black!100] (4,1,3) -- (4,2,3);
\draw[color=black!100] (4,1,3) -- (4,0,3);
\draw[color=black!100] (4,1,3) -- (4,1,4);
\draw[color=black!100] (4,1,3) -- (4,1,2);
\draw[color=black!100] (4,1,4) -- (4,2,4);
\draw[color=black!100] (4,1,4) -- (4,0,4);
\draw[color=black!100] (4,1,4) -- (4,1,5);
\draw[color=black!100] (4,1,4) -- (4,1,3);
\draw[color=black!100] (4,1,5) -- (4,2,5);
\draw[color=black!100] (4,1,5) -- (4,0,5);
\draw[color=black!100] (4,1,5) -- (4,1,4);
\draw[color=black!100] (4,2,1) -- (4,3,1);
\draw[color=black!100] (4,2,1) -- (4,1,1);
\draw[color=black!100] (4,2,1) -- (4,2,2);
\draw[color=black!100] (4,2,2) -- (4,3,2);
\draw[color=black!100] (4,2,2) -- (4,1,2);
\draw[color=black!100] (4,2,2) -- (4,2,3);
\draw[color=black!100] (4,2,2) -- (4,2,1);
\draw[color=black!100] (4,2,3) -- (4,3,3);
\draw[color=black!100] (4,2,3) -- (4,1,3);
\draw[color=black!100] (4,2,3) -- (4,2,4);
\draw[color=black!100] (4,2,3) -- (4,2,2);
\draw[color=black!100] (4,2,4) -- (4,3,4);
\draw[color=black!100] (4,2,4) -- (4,1,4);
\draw[color=black!100] (4,2,4) -- (4,2,5);
\draw[color=black!100] (4,2,4) -- (4,2,3);
\draw[color=black!100] (4,2,5) -- (4,3,5);
\draw[color=black!100] (4,2,5) -- (4,1,5);
\draw[color=black!100] (4,2,5) -- (4,2,4);
\draw[color=black!100] (4,3,1) -- (4,4,1);
\draw[color=black!100] (4,3,1) -- (4,2,1);
\draw[color=black!100] (4,3,1) -- (4,3,2);
\draw[color=black!100] (4,3,2) -- (4,4,2);
\draw[color=black!100] (4,3,2) -- (4,2,2);
\draw[color=black!100] (4,3,2) -- (4,3,3);
\draw[color=black!100] (4,3,2) -- (4,3,1);
\draw[color=black!100] (4,3,3) -- (4,4,3);
\draw[color=black!100] (4,3,3) -- (4,2,3);
\draw[color=black!100] (4,3,3) -- (4,3,4);
\draw[color=black!100] (4,3,3) -- (4,3,2);
\draw[color=black!100] (4,3,4) -- (4,4,4);
\draw[color=black!100] (4,3,4) -- (4,2,4);
\draw[color=black!100] (4,3,4) -- (4,3,5);
\draw[color=black!100] (4,3,4) -- (4,3,3);
\draw[color=black!100] (4,3,5) -- (4,4,5);
\draw[color=black!100] (4,3,5) -- (4,2,5);
\draw[color=black!100] (4,3,5) -- (4,3,4);
\draw[color=black!100] (4,4,1) -- (4,3,1);
\draw[color=black!100] (4,4,1) -- (4,4,2);
\draw[color=black!100] (4,4,2) -- (4,3,2);
\draw[color=black!100] (4,4,2) -- (4,4,3);
\draw[color=black!100] (4,4,2) -- (4,4,1);
\draw[color=black!100] (4,4,3) -- (4,3,3);
\draw[color=black!100] (4,4,3) -- (4,4,4);
\draw[color=black!100] (4,4,3) -- (4,4,2);
\draw[color=black!100] (4,4,4) -- (4,3,4);
\draw[color=black!100] (4,4,4) -- (4,4,5);
\draw[color=black!100] (4,4,4) -- (4,4,3);
\draw[color=black!100] (4,4,5) -- (4,3,5);
\draw[color=black!100] (4,4,5) -- (4,4,4);
\draw (4,0,1) node[tnnode,draw=black!100,text=black!100] {$d$};
\draw (4,0,2) node[tnnode,draw=black!100,text=black!100] {$+$};
\draw (4,0,3) node[tnnode,draw=black!100,text=black!100] {$d$};
\draw (4,0,4) node[tnnode,draw=black!100,text=black!100] {$+$};
\draw (4,0,5) node[tnnode,draw=black!100,text=black!100] {$d$};
\draw (4,1,1) node[tnnode,draw=black!100,text=black!100] {$+$};
\draw (4,1,2) node[tnnode,draw=black!100,text=black!100] {$d$};
\draw (4,1,3) node[tnnode,draw=black!100,text=black!100] {$+$};
\draw (4,1,4) node[tnnode,draw=black!100,text=black!100] {$d$};
\draw (4,1,5) node[tnnode,draw=black!100,text=black!100] {$+$};
\draw (4,2,1) node[tnnode,draw=black!100,text=black!100] {$d$};
\draw (4,2,2) node[tnnode,draw=black!100,text=black!100] {$+$};
\draw (4,2,3) node[tnnode,draw=black!100,text=black!100] {$d$};
\draw (4,2,4) node[tnnode,draw=black!100,text=black!100] {$+$};
\draw (4,2,5) node[tnnode,draw=black!100,text=black!100] {$d$};
\draw (4,3,1) node[tnnode,draw=black!100,text=black!100] {$+$};
\draw (4,3,2) node[tnnode,draw=black!100,text=black!100] {$d$};
\draw (4,3,3) node[tnnode,draw=black!100,text=black!100] {$+$};
\draw (4,3,4) node[tnnode,draw=black!100,text=black!100] {$d$};
\draw (4,3,5) node[tnnode,draw=black!100,text=black!100] {$+$};
\draw (4,4,1) node[tnnode,draw=black!100,text=black!100] {$d$};
\draw (4,4,2) node[tnnode,draw=black!100,text=black!100] {$+$};
\draw (4,4,3) node[tnnode,draw=black!100,text=black!100] {$d$};
\draw (4,4,4) node[tnnode,draw=black!100,text=black!100] {$+$};
\draw (4,4,5) node[tnnode,draw=black!100,text=black!100] {$d$};
\end{tikzpicture}

%% file: figures/gate_splitting_special_case.tex
\begin{tikzpicture}[
square/.style={regular polygon,regular polygon sides=4,inner sep=0,minimum height=22, align=center}]
\draw[draw=none] (0,-2.5) rectangle ++(0,3);
\draw (-2,0) -- (2,0);
\draw[thick,densely dotted] (-2.1,0) -- (-2.27,0);
\draw[thick,densely dotted] (2.1,0) -- (2.27,0);
\draw (-1,-1) -- (-1,1);
\draw (1,-1) -- (1,1);
\draw (-1.7,-0.7) -- (-0.3,0.7);
\draw[thick,densely dotted] (-1.75,-0.75) -- (-1.87,-0.87);
\draw[thick,densely dotted] (-0.25,0.75) -- (-0.13,0.87);
\draw (0.3,-0.7) -- (1.7,0.7);
\draw[thick,densely dotted] (0.25,-0.75) -- (0.13,-0.87);
\draw[thick,densely dotted] (1.75,0.75) -- (1.87,0.87);
\node[square,draw,fill=white] at (-1,0) {+};
\node[square,draw,fill=white] at (0,0) {\eq};
\node[square,draw,fill=white] at (1,0) {+};
\end{tikzpicture}
\begin{tikzpicture}
\draw[draw=none] (-0.6,-2.5) rectangle ++(1.2,3);
\node at (0,0){\large$=$};
\end{tikzpicture}
\begin{tikzpicture}[
square/.style={regular polygon,regular polygon sides=4,inner sep=-10,minimum height=22, align=center}]
\draw (-2,0) -- (-1,0);
\draw (1,0) -- (2,0);
\draw[thick,densely dotted] (-2.1,0) -- (-2.27,0);
\draw[thick,densely dotted] (2.1,0) -- (2.27,0);
\draw (-1,-2.5) -- (-1,1);
\draw (1,-2.5) -- (1,1);
\draw (-1.7,-0.7) -- (-0.3,0.7);
\draw[yshift=-.25cm] (-1,-1.4) -- (1,-1.4);
\draw[thick,densely dotted] (-1.75,-0.75) -- (-1.87,-0.87);
\draw[thick,densely dotted] (-0.25,0.75) -- (-0.13,0.87);
\draw (0.3,-0.7) -- (1.7,0.7);
\draw[thick,densely dotted] (0.25,-0.75) -- (0.13,-0.87);
\draw[thick,densely dotted] (1.75,0.75) -- (1.87,0.87);
\node[square,draw,fill=white] at (-1,0) {+};
\node[square,draw,fill=white] at (1,0) {+};
\begin{scope}[yshift=-0.25cm]
\node[square,draw,fill=white] at (-1,-1.4) {+};
\node[square,draw,fill=white] at (1,-1.4) {+};
\node[square,draw,fill=white] at (0,-1.4) {\eq};
\draw[densely dashed,red,thick, rounded corners=0.25cm] (-1.5,-1.9) rectangle ++(3.0,1.0);
\node at (2.8,-1.4) {\textbf{\color{red}2-qubit gate}};
\end{scope}
\end{tikzpicture}

%% file: figures/snaking.tex
\begin{tikzpicture}[
square/.style={regular polygon,regular polygon sides=4,inner sep=0,minimum height=10, align=center}, scale=1.5]

\draw (-0.2,-0.4) -- (0.6,1.2);
\draw (0.8,-0.4) -- (1.6,1.2);
\draw (1.8,-0.4) -- (2.6,1.2);
\draw[thick,densely dotted] (-0.21,-0.42) -- (-0.26,-0.52);
\draw[thick,densely dotted] (0.79,-0.42) -- (0.74,-0.52);
\draw[thick,densely dotted] (1.79,-0.42) -- (1.74,-0.52);
\draw[thick,densely dotted] (0.61,1.22) -- (0.66,1.32);
\draw[thick,densely dotted] (1.61,1.22) -- (1.66,1.32);
\draw[thick,densely dotted] (2.61,1.22) -- (2.66,1.32);
\draw (-0.5,0) -- (2.5,0);
\draw (-0.1,0.8) -- (2.9,0.8);
\draw[thick,densely dotted] (-0.52,0) -- (-0.64,0);
\draw[thick,densely dotted] (2.52,0) -- (2.64,0);
\draw[thick,densely dotted] (-0.12,0.8) -- (-0.24,0.8);
\draw[thick,densely dotted] (2.92,0.8) -- (3.04,0.8);
\draw (0.0,-0.3) -- (0.0,0.3);
\draw (1.0,-0.3) -- (1.0,0.3);
\draw (2.0,-0.3) -- (2.0,0.3);
\draw (0.4,0.5) -- (0.4,1.1);
\draw (1.4,0.5) -- (1.4,1.1);
\draw (2.4,0.5) -- (2.4,1.1);
\draw[thick,densely dotted] (0.0,-0.32) -- (0.0,-0.44);
\draw[thick,densely dotted] (0.0,0.32) -- (0.0,0.44);
\draw[thick,densely dotted] (1.0,-0.32) -- (1.0,-0.44);
\draw[thick,densely dotted] (1.0,0.32) -- (1.0,0.44);
\draw[thick,densely dotted] (2.0,-0.32) -- (2.0,-0.44);
\draw[thick,densely dotted] (2.0,0.32) -- (2.0,0.44);
\draw[thick,densely dotted] (0.4,0.48) -- (0.4,0.36);
\draw[thick,densely dotted] (0.4,1.12) -- (0.4,1.24);
\draw[thick,densely dotted] (1.4,0.48) -- (1.4,0.36);
\draw[thick,densely dotted] (1.4,1.12) -- (1.4,1.24);
\draw[thick,densely dotted] (2.4,0.48) -- (2.4,0.36);
\draw[thick,densely dotted] (2.4,1.12) -- (2.4,1.24);

\node[square,draw,fill=white] (A) at (0,0) {};
\node[square,draw,fill=white] (B) at (1,0) {};
\node[square,draw,fill=white] (C) at (2,0) {};
\node[square,draw,fill=white] (D) at (0.4,0.8) {};
\node[square,draw,fill=white] (E) at (1.4,0.8) {};
\node[square,draw,fill=white] (F) at (2.4,0.8) {};

\node[square,draw,fill=white,thick] (eq) at (0.7,0.4) {\tiny =};
\draw[thick] (A) -- (eq);
\draw[thick] (B) -- (eq);
\draw[thick] (F) -- (eq);

\draw [-stealth](3.3,0.4) -- (3.8,0.4);
\end{tikzpicture}
\begin{tikzpicture}[
square/.style={regular polygon,regular polygon sides=4,inner sep=0,minimum height=10, align=center}, scale=1.5]

\draw (-0.2,-0.4) -- (0.6,1.2);
\draw (0.8,-0.4) -- (1.6,1.2);
\draw (1.8,-0.4) -- (2.6,1.2);
\draw[thick,densely dotted] (-0.21,-0.42) -- (-0.26,-0.52);
\draw[thick,densely dotted] (0.79,-0.42) -- (0.74,-0.52);
\draw[thick,densely dotted] (1.79,-0.42) -- (1.74,-0.52);
\draw[thick,densely dotted] (0.61,1.22) -- (0.66,1.32);
\draw[thick,densely dotted] (1.61,1.22) -- (1.66,1.32);
\draw[thick,densely dotted] (2.61,1.22) -- (2.66,1.32);
\draw (-0.5,0) -- (2.5,0);
\draw (-0.1,0.8) -- (2.9,0.8);
\draw[thick,densely dotted] (-0.52,0) -- (-0.64,0);
\draw[thick,densely dotted] (2.52,0) -- (2.64,0);
\draw[thick,densely dotted] (-0.12,0.8) -- (-0.24,0.8);
\draw[thick,densely dotted] (2.92,0.8) -- (3.04,0.8);
\draw (0.0,-0.3) -- (0.0,0.3);
\draw (1.0,-0.3) -- (1.0,0.3);
\draw (2.0,-0.3) -- (2.0,0.3);
\draw (0.4,0.5) -- (0.4,1.1);
\draw (1.4,0.5) -- (1.4,1.1);
\draw (2.4,0.5) -- (2.4,1.1);
\draw[thick,densely dotted] (0.0,-0.32) -- (0.0,-0.44);
\draw[thick,densely dotted] (0.0,0.32) -- (0.0,0.44);
\draw[thick,densely dotted] (1.0,-0.32) -- (1.0,-0.44);
\draw[thick,densely dotted] (1.0,0.32) -- (1.0,0.44);
\draw[thick,densely dotted] (2.0,-0.32) -- (2.0,-0.44);
\draw[thick,densely dotted] (2.0,0.32) -- (2.0,0.44);
\draw[thick,densely dotted] (0.4,0.48) -- (0.4,0.36);
\draw[thick,densely dotted] (0.4,1.12) -- (0.4,1.24);
\draw[thick,densely dotted] (1.4,0.48) -- (1.4,0.36);
\draw[thick,densely dotted] (1.4,1.12) -- (1.4,1.24);
\draw[thick,densely dotted] (2.4,0.48) -- (2.4,0.36);
\draw[thick,densely dotted] (2.4,1.12) -- (2.4,1.24);

\node[square,draw,fill=white] (A) at (0,0) {};
\node[square,draw,fill=white,thick] (Aeq) at (0.25,0.15) {\tiny =};
\node[square,draw,fill=white] (B) at (1,0) {};
\node[square,draw,fill=white,thick] (Beq) at (1.25,0.15) {\tiny =};
\node[square,draw,fill=white] (C) at (2,0) {};
\node[square,draw,fill=white,thick] (Ceq) at (2.25,0.15) {\tiny =};
\node[square,draw,fill=white] (D) at (0.4,0.8) {};
\node[square,draw,fill=white] (E) at (1.4,0.8) {};
\node[square,draw,fill=white] (F) at (2.4,0.8) {};
\node[square,draw,fill=white,thick] (Feq) at (2.65,0.95) {\tiny =};

\draw[thick] (A) -- (Aeq);
\draw[thick] (B) -- (Beq);
\draw[thick] (F) -- (Feq);
\draw[thick] (Aeq) -- (Beq);
\draw[thick] (Beq) -- (Ceq);
\draw[thick] (Ceq) -- (Feq);
\draw[rotate around={300:(0.125,0.075)},red,thick,
dash pattern=on 2pt off 1pt] (0.125,0.075) ellipse (0.175 and 0.35);
\draw[rotate around={300:(1.125,0.075)},red,thick,
dash pattern=on 2pt off 1pt] (1.125,0.075) ellipse (0.175 and 0.35);
\draw[rotate around={300:(2.125,0.075)},red,thick,
dash pattern=on 2pt off 1pt] (2.125,0.075) ellipse (0.175 and 0.35);
\draw[rotate around={300:(2.525,0.875)},red,thick,
dash pattern=on 2pt off 1pt] (2.525,0.875) ellipse (0.175 and 0.35);
\end{tikzpicture}